\documentclass{WileyMSP-template}
\usepackage{xcolor}
\usepackage{amsmath}
\usepackage{amssymb}
\usepackage[colorlinks,linkcolor=blue,citecolor=blue,urlcolor=blue]{hyperref}

\let\oldAA\AA
\renewcommand{\AA}{\text{\normalfont\oldAA}}
\usepackage{lineno}

\begin{document}


\pagestyle{fancy}
\rhead{\includegraphics[width=2.5cm]{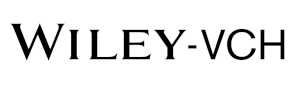}}

\title{Observation of multi-directional energy transfer\\
in a hybrid plasmonic-excitonic nanostructure}

\maketitle


\author{Tommaso Pincelli* $^\dagger$}
\author{Thomas Vasileiadis $^\dagger$}
\author{Shuo Dong}
\author{Samuel Beaulieu}
\author{Maciej Dendzik}
\author{Daniela Zahn}
\author{Sang-Eun Lee}
\author{Hélène Seiler}
\author{Yinpeng Qi}
\author{R.Patrick Xian}
\author{Julian Maklar}
\author{Emerson Coy}
\author{Niclas S. Mueller}
\author{Yu Okamura}
\author{Stephanie Reich}
\author{Martin Wolf}
\author{Laurenz Rettig}
\author{Ralph Ernstorfer*}\\
$^\dagger$ These authors contributed equally.


\dedication{}

\begin{affiliations}
Dr. T. Pincelli, Dr. T. Vasileiadis, Dr. S. Dong, Dr. S. Beaulieu, Dr. M. Dendzik, Dr. D. Zahn, S.-E. Lee, Prof. H. Seiler, Dr. Y. Qi, Dr. R. P. Xian, J. Maklar, Prof. M. Wolf, Dr. L. Rettig, Prof. Ernstorfer\\
Fritz-Haber-Institut der Max-Planck-Gesellschaft, Faradayweg 4-6, 14195 Berlin, Germany\\
Email Address: pincelli@fhi-berlin.mpg.de, ernstorfer@tu-berlin.de\\

Prof. H. Seiler, Dr. N. S. Mueller, Y. Okamura, Prof. S. Reich\\
Freie Universität Berlin, Arnimallee 14, 14195 Berlin, Germany.\\

Dr. T. Pincelli, Prof. R. Ernstorfer\\
Institut für Optik und Atomare Physik, Technische Universität Berlin, Straße des 17.~Juni 135, 10623 Berlin, Germany\\

Dr. T. Vasileiadis\\
Faculty of Physics, Adam Mickiewicz University, Uniwersytetu Poznanskiego 2, 61-614 Poznan, Poland\\

Dr. S. Beaulieu\\
Université de Bordeaux - CNRS - CEA, CELIA, UMR5107, F33405, Talence, France.\\

Dr. M. Dendzik\\
Department of Applied Physics, KTH Royal Institute of Technology, Hannes Alfvéns väg 12, 114 19 Stockholm, Sweden.\\

Dr. Y. Qi\\
Center for Ultrafast Science and Technology, School of Physics and Astronomy, Shanghai Jiao Tong University, 200240 Shanghai, China.\\

Dr. R. P. Xian\\
Department of Statistical Sciences, University of Toronto, 700 University Avenue, Toronto, M5G 1Z5, Canada.\\

Dr. E. Coy\\
NanoBioMedical Centre, Adam Mickiewicz University, ul. Wszechnicy Piastowskiej 3, PL 61614 Poznań, Poland.\\

Dr. N. S. Mueller\\
NanoPhotonics Centre, Cavendish Laboratory, Department of Physics, University of Cambridge, JJ Thomson Avenue, Cambridge CB30HE, United Kingdom.\\
\end{affiliations}



\begin{abstract}
\justifying
Hybrid plasmonic devices involve a nanostructured metal supporting localized surface plasmons to amplify light-matter interaction, and a non-plasmonic material to functionalize charge excitations. Application-relevant epitaxial heterostructures, however, give rise to ballistic ultrafast dynamics that challenge the conventional semiclassical understanding of unidirectional nanometal-to-substrate energy transfer.
We study epitaxial Au nanoislands on WSe$_2$ with time- and angle-resolved photoemission spectroscopy and femtosecond electron diffraction: this combination of techniques resolves material, energy and momentum  of charge-carriers and phonons excited in the heterostructure.
We observe a strong non-linear plasmon-exciton interaction that transfers the energy of sub-bandgap photons very efficiently to the semiconductor, leaving the metal cold until non-radiative exciton recombination heats the nanoparticles on hundreds of femtoseconds timescales. Our results resolve a multi-directional energy exchange on timescales shorter than the electronic thermalization of the nanometal. Electron-phonon coupling and diffusive charge-transfer determine the subsequent energy flow.
This complex dynamics opens perspectives for optoelectronic and photocatalytic applications, while providing a constraining experimental testbed for state-of-the-art modelling.

\end{abstract}


\justifying
\section{Introduction}
Irradiation of nanometals with light drives collective oscillations of charge-carriers (plasmons) and light localization beyond the diffraction limit in plasmonic near-fields. The energy of plasmons dissipates within tens of femtoseconds, either radiatively by photon emission, or in electron-hole excitations, producing non-equilibrium carrier distributions. 

In recent years, the focus of plasmonics is oriented towards plasmonic energy harvesting~\cite{Moskovits2015,brongersma_plasmon-induced_2015,Linic2021}. The nascent field of hybrid plasmonics seeks to interface metal nanostructures with other materials, and in particular semiconductors, which convert plasmons to electronic excitations with impactful applications. Hybrid plasmonics devices are useful in light-harvesting, photochemistry, photocatalysis, photodetectors and single-molecule detectors~\cite{atwater_plasmonics_2010,clavero_plasmon-induced_2014,brongersma_plasmon-induced_2015,furube_insight_2017,langer_present_2020}. For these applications, radiative losses are suppressed, e.g., by minimizing the metallic nanoparticles' volume~\cite{Jain2006}, while hot-carrier injection is maximized by creating a strong exciton-plasmon interaction. This is achieved when the plasmonic hot carriers ejection is more efficient than internal thermalization by electron-phonon coupling.
 
The challenges in the microscopic description of hybrid plasmonic systems stem from the strong inhomogeneity of the hot-carrier distributions both in real space and, for the case of crystals with well-defined Bloch states, also in reciprocal space. In these conditions, the assumptions of homogeneous hot carrier generation and instant thermalization (that proved effective for larger nanoparticles in colloidal suspensions) are no longer applicable: geometry-assisted intraband transitions dominate plasmon decoherence dynamics enhancing the hot carrier distribution at the surface~\cite{Khurgin2015,Boerigter2016}, narrow gaps between the nanoparticles generate regions of enhanced polarization of the substrate~\cite{Hartland2017}, and unoccupied states in the semiconductor offer high-energy excitation transitions across the interface at selected momentum matched locations~\cite{tan_plasmonic_2017,Tagliabue_2020}. Furthermore, the coupling of the electronic system with phononic excitations, and the subsequent energy flow leading back to thermodynamic equilibrium, are critical in determining the subsequent functionalities achievable by devices based on the hybrid interface~\cite{Brown2017}. 

We focus on achieving a fundamental understanding of how electron-plasmon interaction influences the interfacial energy flow in a rapidly emerging class of heterostructures formed by noble metals interfaced with transition metal dichalcogenides. The 2D van der Waals crystals, with their long lived and strongly bound excitons, promise a unique playground for hybrid plasmonics, and have been proven to exhibit strong exciton-plasmon coupling up to room temperature~\cite{Kleemann2017,Yan2020}. 
The TMD-noble metal interface has been realized both in configurations where the TMD has a nanoscale structure~\cite{Cabo_2015_MoS2onAu, Vogelsang2021, Xu2020} and in devices where the 3D metal is laterally confined~\cite{Li_2020, shan_direct_2019, Au_TMDC_ACSphotonics_2020, Yan2020, Lin2022, Kleemann2017}. Nano-TMDs on extended Au present fascinating possibilities for the control of surface plasmons owing to their large dielectric functions~\cite{Vogelsang2021} and have enjoyed a significant effort devoted to understanding the microscopic mechanisms behind interfacial dynamics and band alignment~\cite{Cabo_2015_MoS2onAu, Dendzik2017, Hinsche2017, Xu2020}. We focus, instead, on nano-structured Au on extended TMD, which has been recently employed to realize several devices with advanced plasmonic functionality~\cite{Li_2020, Lin2022} or strong coupling~\cite{Kleemann2017, shan_direct_2019, Au_TMDC_ACSphotonics_2020, Yan2020}, but whose investigation of microscopic mechanisms has been less extensive.

Satisfactory modelling of microscopic charge-transfer mechanisms has been recently achieved with \textit{ab-initio} calculations that go beyond the simple jellium model~\cite{Varas_2015, Tagliabue_2020}, but disentangling the same processes experimentally is challenging, as it requires monitoring several microscopic subsystems and their couplings at femtosecond timescales, as shown in Figure~\ref{fig:fig_1}~a. The vast majority of experimental studies employed time-resolved optical spectroscopies~\cite{shan_direct_2019,Au_TMDC_ACSphotonics_2020}, with techniques that have limited access to optically dark excitations and to the details of quasiparticle scattering pathways in momentum space. We propose an approach that offers the opportunity to ground the study of plasmonic dynamics with an unprecedented detail of physical evidence, resolving the dynamics of the electronic states in momentum space and the coupling of charge excitations to phonons: this provides a direct response to the needs of hybrid plasmonics as foreseen by Linic et al.\cite{Linic2021}.

Here, we study a heterostructure of Au nanoislands on WSe$_{2}$ (Fig.~\ref{fig:fig_1}~a) with time- and angle-resolved photoemission spectroscopy (tr-ARPES) that gives access to equilibrium and excited electronic states with momentum resolution~\cite{Cabo_2015_MoS2onAu,Bertoni_2016,Nicholson_2018} (Fig.~\ref{fig:fig_1}~b), captures excitons in the TMD~\cite{Dong_2021} and detects the dynamic hot-carrier distributions in the nanometal~\cite{Sygletou2021}. To fully unfold the dynamics of the system we investigate also the complementary subsystem, the lattice, by femtosecond electron diffraction (FED)~\cite{Waldecker_2015,waldecker_time-domain_2015,Waldecker_PRL_2017}(Fig.~\ref{fig:fig_1}~c), determining the coupling of electronic excitations to phononic states. 

Further support is provided by electronic structure calculations using density functional theory and finite elements calculations to investigate the distribution of interfacial fields. With this toolset we show that strong exciton-plasmon interactions can produce multi-directional charge- and energy-flow between metal and semiconductor at extremely short timescales, a picture rather different from what is expected in traditional plasmonic approaches and semiclassical models both at the femtosecond and picosecond timescale. By determining the origins and timescales of such transient energy transfers, our results provide important insight for the design of a wide set of hybrid heterostructures that are object of very active investigation~\cite{Fast_epitaxy_1999,Fast_epitaxy_2018,Fast_epitaxy_2019}, as well as an accurate and constraining experimental test for the advanced theoretical models being developed for hybrid plasmonics. 

\section{Results}

\begin{figure}[t!]
\centering\includegraphics[width=130mm]{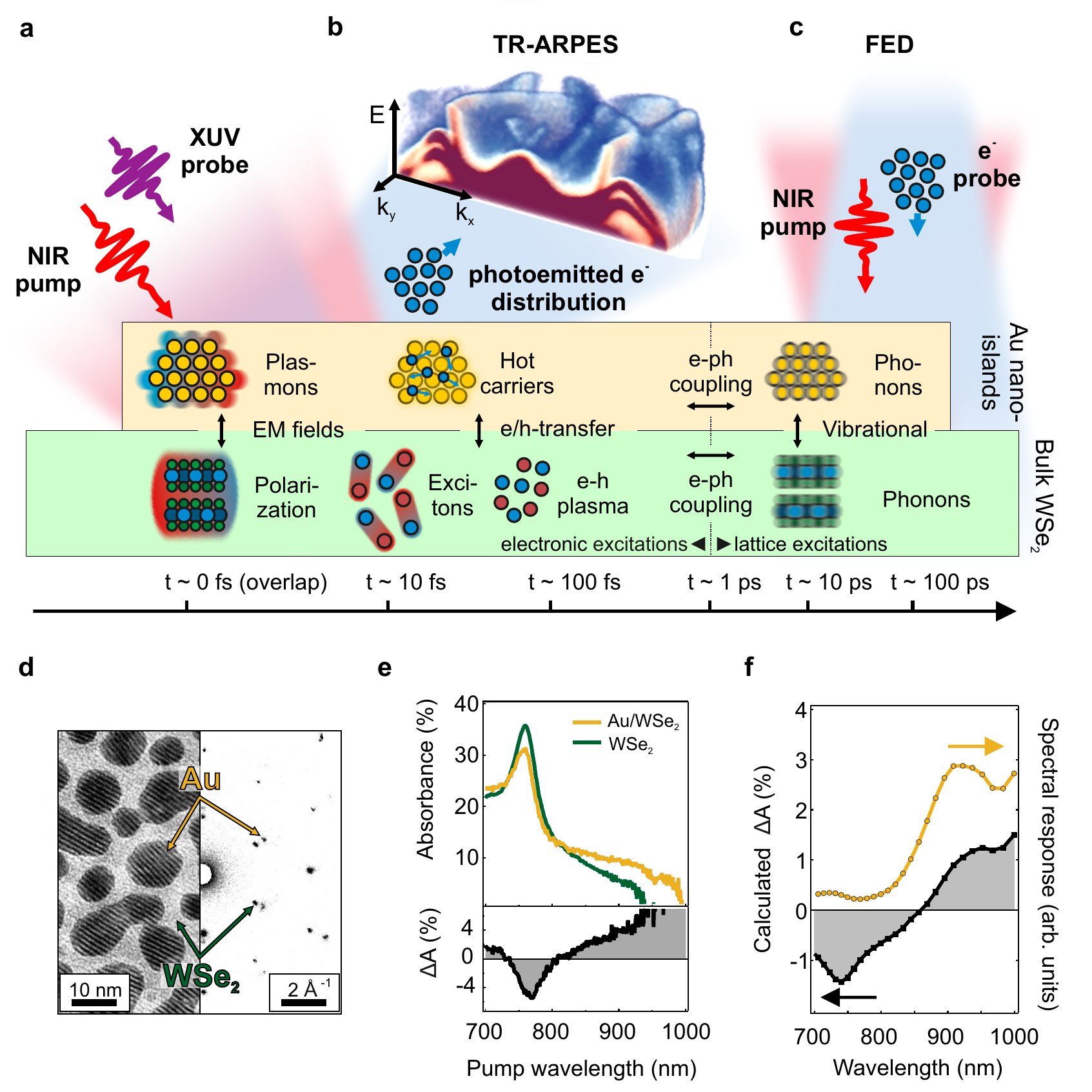}
\caption{\footnotesize \textbf{Subsystem- and material-resolved study of ultrafast energy flow in a nanometal-semiconductor heterostructure.} \textbf{(a)} Schematic illustration of the techniques employed to probe different microscopic subsystems and their couplings in the heterostructure. Optical excitation by a pump pulse generates plasmons and hot carriers in the nanometal, and excitons in the semiconductor, probed by tr-ARPES (schematically represented in panel b). In both materials the lattice degrees of freedom (phonons) are excited by electron-phonon coupling, probed by FED (schematically represented in c and the right part of panel f). Interfacial interactions involve: electromagnetic fields (LSP and interfacial potentials), charge and energy transfer (Meitner-Auger, Föster or Dexter coupling) and vibrational coupling. \textbf{(b)} Cut-out (for k$_y>$0) of the 3D ARPES signal I(k$_x$,k$_y$,E) from the heterostructure surface with 21.7 eV XUV excitation. In a tr-ARPES experiment, one of such volumetric datasets is acquired for each pump-probe delay.  \textbf{(c)} Scheme of a transmission FED experiment. \textbf{(d)} Electron microscopy (left) and electron diffraction (right) of epitaxial Au nanoislands (dark in the micrograph) on single-crystalline multilayer flakes of WSe$_2$. In a FED experiment, performed in transmission  geometry, one of such diffractograms is acquired for every pump-probe delay. \textbf{(e)} Upper panel: light absorption spectra of pure (green line) and Au-decorated WSe$_2$ (yellow line). Bottom panel: wavelength-dependent relative change of the absorption ($\Delta$A) due to Au decoration. \textbf{(f)} Black squares: Calculated relative change of absorption using finite difference modelling (see Methods). Golden circles: spectral response of the electric field integral.}
\label{fig:fig_1}
\end{figure}

The Au nanoislands grow epitaxially on bulk WSe$_2$ with a random distribution of sizes and shapes, as displayed by TEM micrography in Fig.~\ref{fig:fig_1}~d. The average lateral size and thickness are 10 nm and 2 nm respectively (see SI Sect. 1). Optical absorption spectroscopy of bare and Au-decorated WSe$_2$ shows a significant modification of the absorbance (Fig.~\ref{fig:fig_1}~e): a suppression of the WSe$_2$ A-exciton peak is accompanied by an increased absorption at $\lambda \geq 800$ nm which, as we will show, arises from localized, plasmon-assisted photoabsorption in the nanoparticle array. The stochastic distribution of sizes and shapes results in a broadly varying enhancement of the absorption within the bandgap of the semiconductor, rather than peaked resonances~\cite{Sun_2012}. 
In the following, we demonstrate that Au decoration tailors the pump electric field at the surface, producing a twofold effect: generation of localized surface plasmons (LSP) on the nanoislands, and strong field enhancement at the uncovered WSe$_2$ surface. 

To gain insight on the optical response, we perform finite-element-method (FEM) calculations (see Methods and SI Sect. 3). We model the nanoparticles by vectorizing a part of the micrograph in Fig.~\ref{fig:fig_1}~d, to reproduce a realistic arrangement of shapes. We obtain the far-field absorbance variation due to the presence of Au decoration, which broadly matches the spectral distribution of the experimental result (black curves in Fig.~\ref{fig:fig_1}~e and f). Discrepancies in amplitude and spectral distributions can be explained with the experimental uncertainty on the effective thickness of the individual sample, combined with the use of reference optical spectra and the limited size of the calculation (see Experimental Section). The FEM calculations enable estimating the contribution of plasmon-induced hot-carriers to the increased absorbance (yellow circles in Fig.~\ref{fig:fig_1}~f, see Methods). The total absorbance variation (black curve) is adequately represented at photon energies smaller than the WSe$_2$ bandgap, thus confirming that, in this spectral range, photoabsorption in Au is dominated by plasmonic generation of injectable hot-carriers~\cite{Zheng_2015,Manjavacas_2014,Govorov_2013}. Since the nanoislands thickness is below the inelastic mean free path (IMFP) of hot carriers in Au~\cite{Brown_2016b} and the out-of-plane momentum matching constraints are relaxed due to strong vertical confinement, virtually all the plasmon-generated hot-carriers energetically above the Schottky barrier (SB) are available for injection. 

For high peak power excitation, easily achieved with ultrashort light pulses, absorption also involves non-linear multiphoton processes. Close examination of the electric fields (see SI Sect. 2) shows a strong field enhancement of up to 30 times in the semiconductor near the nanoislands edges, which can induce multiphoton absorption. The amplitude and depth penetration of the field enhancement in WSe$_2$ increase as the photon energy becomes smaller than the A-exciton energy. 
In this spectral range, WSe$_2$ becomes more transparent and the plasmonic fields propagate further below the surface. Therefore, at high optical excitation intensities and long ($\geq 850$~nm) wavelengths, we expect the photoabsorption in WSe$_2$ to occur predominantly by multiphoton processes.

From the volumetric ARPES data displayed in Fig.~\ref{fig:fig_1}~b, we extract isoenergy maps of the momentum distribution. At the Fermi energy, the hexagonal \textit{sp}-band and Shockley surface state form the Fermi surface of Au (111) (Fig.~\ref{fig:fig_2}~a, top panel). At E-E$_F$=-0.8 eV (Fig.~\ref{fig:fig_2}~a, bottom panel) we observe that the ARPES signal appears as the superposition of Au (111) and WSe$_2$ bandstructures, with the Au \textit{sp}-band surrounded by the hexagonal arrangement of WSe$_2$ valence band maxima (VBM). By selecting a momentum direction ($\overline{\Gamma}-\overline{K}$ in the 2D Brillouin zone of WSe$_2$, black line in Fig.~\ref{fig:fig_2}~a), we extract a momentum-energy map, and compare with ab-initio calculations (see SI Sect. 3) employing density functional theory (DFT) as overlaid on the data in Fig.~\ref{fig:fig_2}~b. 
While the calculations have been performed for the two separate materials, the good agreement between DFT and experiment suggests a weak hybridization across the van der Waals gap (see SI Sect. 4). 

The ARPES data combined with further core-level photoemission experiments reveal the details of band alignment (Fig.~\ref{fig:fig_2}~c): considering the Schottky-Mott theory of contact potential, the Fermi level would be expected to be energetically near the valence band maximum. However, owing to the work function reduction observed in Au nanoparticles~\cite{Zhang_2015} and interfacial dipoles formation, the Fermi level is closer to the conduction band minimum of WSe$_2$, with Schottky barrier $\Phi_e=0.470\pm0.005$ eV for electrons (ESB) and $\Phi_h=1.000\pm0.005$ eV for holes (HSB)~\cite{Smyth_2017}  (see SI Sect. 5). With this configuration, interfacial band bending is strongly suppressed, although a potential well of approximately 170 meV still exists, in the proximity of the nanoparticle interface.

\subsection{Electron dynamics}

\begin{figure}[t!]
\centering\includegraphics[width=130mm]{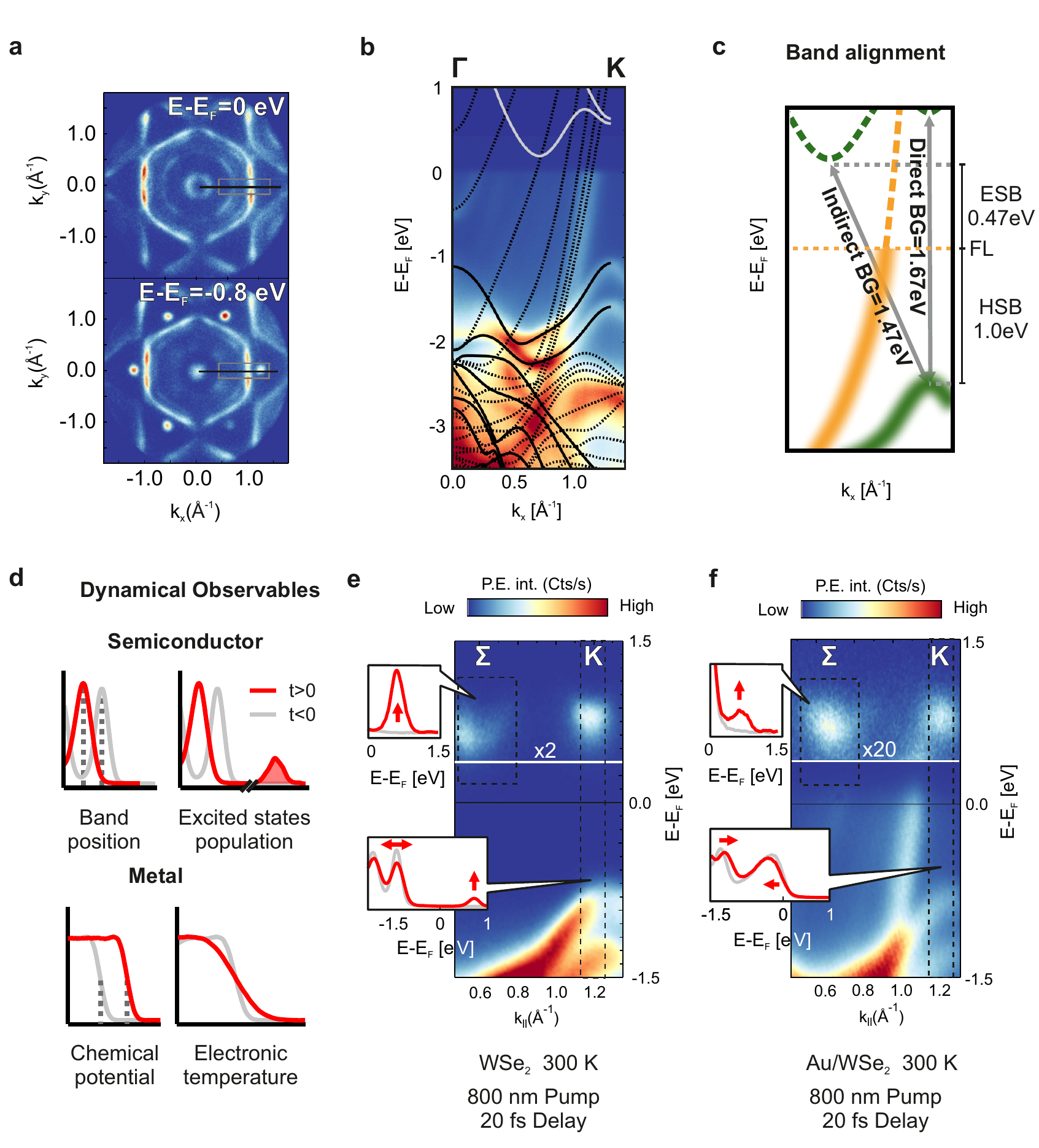}
\caption{\footnotesize \textbf{Electronic structure of Au/WSe$_2$.} \textbf{(a)} Constant energy cuts of Au/WSe$_2$ in the first Brillouin zone. Top panel: Fermi surface E-E$_F$=0 eV, with hexagonal sp-band and surface state (close to the $\Gamma$ point), characteristic of the Au(111) facet. Bottom panel: isoenergy surface at E-E$_F$=-0.8 eV, showing the VBM at the K-points of WSe$_2$ (hexagonal array of dots). The black line (grey rectangle) shows the energy-momentum cut probed in panel b (panel f). \textbf{(b)} $\Gamma$-K momentum-energy cut of Au/WSe$_2$. The solid black and white curves show the valence and conduction bands of WSe$_2$ as derived by DFT. The dashed curves show the DFT-derived metallic states of Au.  \textbf{(c)} Schematic picture of the band alignment: at the WSe$_2$ K points, the direct band gap (DBG) is 1.67 eV, larger than the indirect bandgap (IBG) at 1.47 eV. Due to nanostructuring, the Fermi Level (FL) position is closer to the conduction band than expected from the Schottky-Mott limit, with Electron and Hole Schottky Barriers (ESB and HSB) of 0.47 eV and 1 eV, respectively. \textbf{(d)} Schematic summarizing the observables employed to probe the dynamics of the heterostructure before (grey) and after (red) excitation. \textbf{(e)} Momentum-energy cut for the bare WSe$_2$ surface, acquired at +20 fs delay. The intensity above the white horizontal line (E-E$_F$=+300 meV) has been multiplied by 2. Both K and $\Sigma$ valleys are populated. The insets show EDCs evolution within the dashed regions. \textbf{(f)} Same as (e), but for the Au/WSe$_2$ heterostructure, at +20 fs. The intensity above the white horizontal line (E-E$_F$=+300 meV) has been multiplied by 20.}
\label{fig:fig_2}
\end{figure}

To explore the charge dynamics during and right after the plasmonic excitation, we perform time-resolved ARPES experiments. Four observables, sketched in Fig.~\ref{fig:fig_2}~d, allow us to distinguish the evolution of each material. In the semiconductor, we detect both the occupied band position and the photoexcited population in the conduction band. 
The time resolved experimental technique adds an additional dimension to the data and allows to capture transient population and scattering dynamics~\cite{Bertoni_2016,Dong_2021}. 
In Au, we can track the dynamical evolution of the chemical potential, i.e.~the energy position of the Fermi distribution center, and the electronic temperature. The main features involved in the dynamics are contained in a single momentum-energy cut (grey squares in Fig.~\ref{fig:fig_2}~a).

\begin{figure}[t!]
\centering\includegraphics[width=130mm]{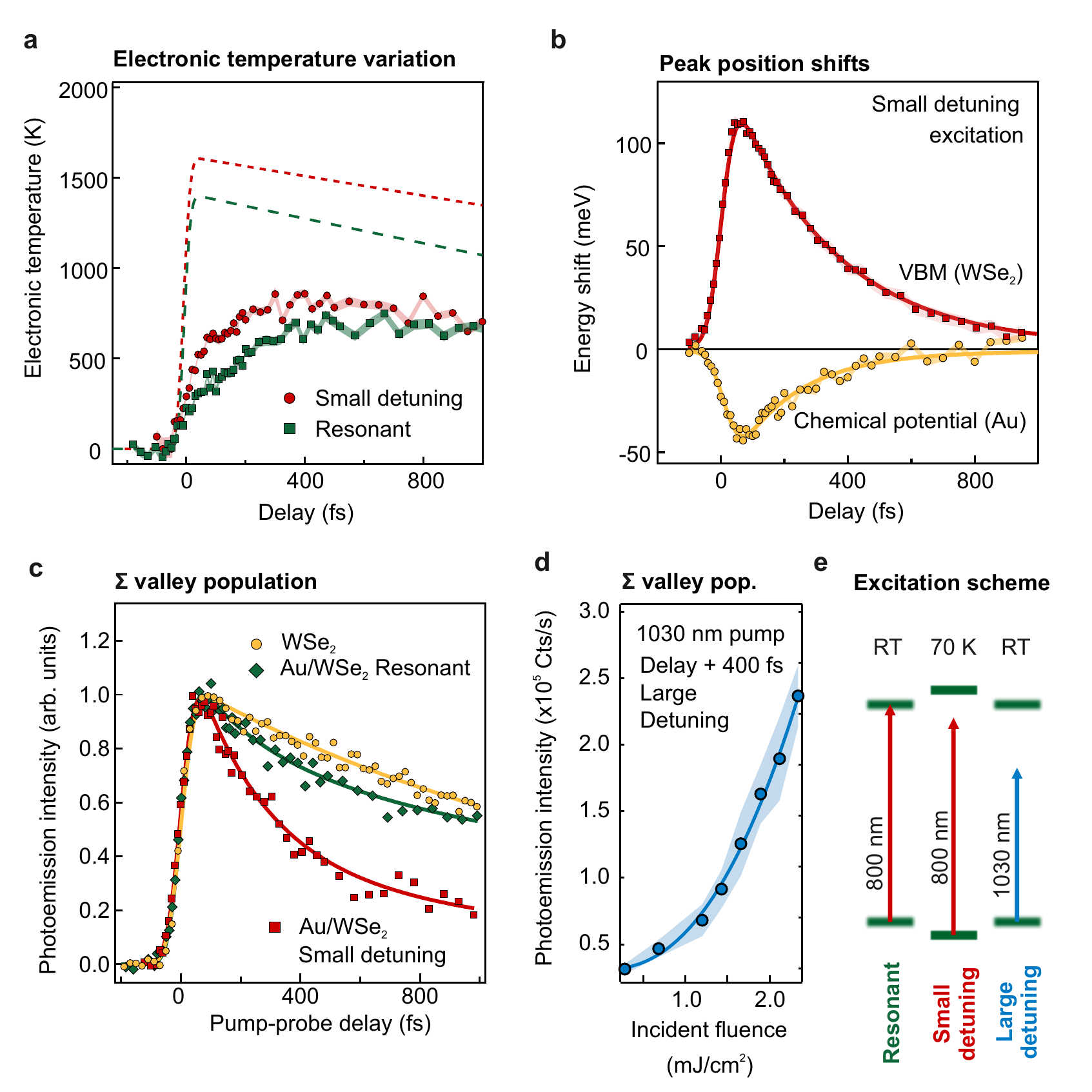}
\caption{\footnotesize \textbf{Multi-observable electronic dynamics in Au/WSe$_2$.} \textbf{(a)} Electronic temperature as extracted from Fermi function fitting of the data, compared with 2TM prediction (dashed lines), using the experimental absorbed energy (from Fig.~\ref{fig:fig_1}~e) and temperature-dependent experimental values of electron-phonon coupling constant, electronic and lattice heat capacities for nanostructured Au. \textbf{(b)} Dynamic shifting of Au Fermi edge (circles) and WSe$_2$ VBM (squares). Solid lines are fits of an exponential convolved with the instrument response function (IRF). The exponential decay characteristic times are $251 \pm 24$ fs (Au) and $318 \pm 7$ fs (WSe$_2$) respectively. \textbf{(c)} Population dynamics in the $\Sigma$ valley for different experimental settings. For bare WSe$_2$, the solid line is an exponential fit convolved with the IRF, giving a decay time of $1.5\pm0.08$ ps. For the heterostructure, the solid lines are fits with double exponential decay convolved with the IRF. The second exponential is fixed at 1.5 ps, while the first is $240\pm30$ fs in both cases. The relative amplitude of the first decay changes from $36\pm8$ \% to $72\pm2$ \% when going from 300 K to 70 K.   \textbf{(d)} Fluence dependence of the $\Sigma$ valley population at long delays vs fluence for 1030 nm photoexcitation. The solid line is a power law fit resulting in an exponent of $2.3 \pm 0.3$. Shaded areas show the experimental uncertainty discussed in the SI. \textbf{(e)} Schematic showing the excitation schemes employed. At room temperature, $\lambda$=800 nm is quasi-resonant with the first excitonic transition. At 70 K, the same wavelength is slightly out of resonance, in the small detunig regime. At room temperature, $\lambda$=1030 nm is strongly detuned, highlighting plasmonic nonlinear effects.}
\label{fig:fig_3}
\end{figure}

In bare WSe$_2$ at room temperature, 800 nm pumping excites resonantly the A-exciton ~\cite{Dong_2021}, thanks to the large bandwidth of 40 fs pulses. The excitons, observable as a transient population at the K-points (Fig.~\ref{fig:fig_2}~e), scatter rapidly ($18\pm4$ fs~\cite{Bertoni_2016,Dong_2021}) to momentum-indirect states, with electrons populating the conduction band minimum ($\Sigma$ valley) and holes occupying local valence band maxima (K or $\Gamma$ points). The lifetime of indirect excitons is long owing to the suppressed radiative recombination probability arising from momentum mismatch~\cite{Wang_2018, Selig_2018, Dong_2021}. We observe a bi-exponential population decay, with the shortest lifetime of $1.5\pm0.1$ ps (see SI Sect. 6) likely determined by hot exciton diffusion in the bulk and surface defect recombination~\cite{Massicotte_2016}. 
In the heterostructure, 800 nm pumping at room temperature produces excited carriers in both Au and WSe$_2$ (Fig.~\ref{fig:fig_2}~f, see also SI Fig. S4). To disentangle the contributions, we perform two measurements: one at room temperature, and another after cooling the heterostructure to 70 K (see also Fig.~\ref{fig:fig_3}~e). 
The reduction of temperature in this system produces an increase of the WSe$_2$ bandgap (in optical measurements: $\approx$60~meV~\cite{Arora_2015}), tuning the exciton resonance out of the pump bandwidth, which we will consider as the small detuning regime.

Fig.~\ref{fig:fig_3}~a reports the electronic temperature (T$_e$) dynamics of Au obtained by fitting the data to Fermi-Dirac distributions (see SI Sect. 11). Compared to the well-known electron-lattice equilibration dynamics of bulk Au (see SI Sect. 12, shown in Fig.~\ref{fig:fig_3} a as a result of two-temperature model), the experimental T$_e$ of the heterostructure rises on a longer timescale and to a lesser degree. This suggests efficient charge-transfer from the Au nanoislands to WSe$_2$, to such an extent that Au hot carriers are transferred before the non-equilibrium distribution can thermalize, a signature of strong exciton-plasmon interaction. The observation of plasmon-induced hot-carrier transfer is further supported by tracking the position of WSe$_2$ VBM and the chemical potential in the small detuning case (Fig.~\ref{fig:fig_3}~b). The two features shift simultaneously in opposite directions, indicating that unbalanced amounts of charge are being transferred across the interface. These observations match optical measurements and FEM calculations, where we highlighted the relevance of LSP-induced injectable hot carriers in the absorption spectrum of the heterostructure (see SI Sect. 8 and 9).

The recovery dynamics in Fig.~\ref{fig:fig_3}~b reveal the mechanisms immediately subsequent the charge-transfer process: an interfacial electrical field is generated, and the increase of the SB combines with hot-carrier relaxation to stop the hot-carrier injection. The injected carriers, after losing energy to the lattice of WSe$_2$, are rapidly back-injected to reach charge compensation, on a timescale of 250-300 fs. The charges re-injected in Au lose energy, producing the slow rise in electronic temperature observed in Fig.~\ref{fig:fig_3}~a. A clear indication of the back-transfer process can be seen in Fig.~\ref{fig:fig_3}~c, where the population dynamics of the $\Sigma$ valley is displayed for the three cases of the bare WSe$_2$, and the heterostructure under resonant exitation and in the small detuning regime. The decay of the population becomes strikingly faster in the heterostructure with small detuning as back-injection offers a rapid channel for charge compensation. We find the decay time of the back-injected population to be $240\pm30$~fs by constrained multi-exponential fitting (see SI Sect. 7). 

In Fig.~\ref{fig:fig_3}~c, the presence of a slow-decaying exciton population is evident at longer delays for the heterostructure. Its relative fraction changes from $70\pm10$ \% to $30\pm10$ \% when the excitation falls out of resonance with the A-exciton. This is the population of charge-neutral, thermalized excitons. They are formed by direct excitation with resonant pumping, and by plasmon-enhanced multiphoton absorption when the excitation is non-resonant. To support this interpretation, we measured the fluence dependence of the $\Sigma$-valley population at long (+400 fs) pump-probe delays in the strong detuning regime (1030 nm pumping at room temperature), reported in Fig.~\ref{fig:fig_3}~d (blue circles). As demonstrated by the finite element calculations (see SI Sect. 2) and as expected from standard nonlinear optics consideration, long wavelengths and high fluences maximize multiphoton absorption in WSe$_2$. Fig.~\ref{fig:fig_3}~e reports all the excitation schemes employed in the experiment. The population scales with fluence following a power law of exponent $2.3\pm0.3$, indicating a two-photon absorption process (see SI Sect. 7).
The thermalized excitons give rise to a long-timescale dynamics as they diffuse towards the nanoparticles and recombine by transferring carriers to Au, discussed in the following.

\subsection{Lattice dynamics}

The vibrational response of the heterostructure is probed by FED~\cite{Waldecker_2015}. Following optical excitation of the electrons, electron-phonon coupling increases the vibrational energy content, leading to the main quantity extracted with FED: the change of the atomic mean-squared-displacement (MSD) $\Delta\langle u^2\rangle$. 
 
\begin{figure}[t!]
\centering\includegraphics[width=86 mm]{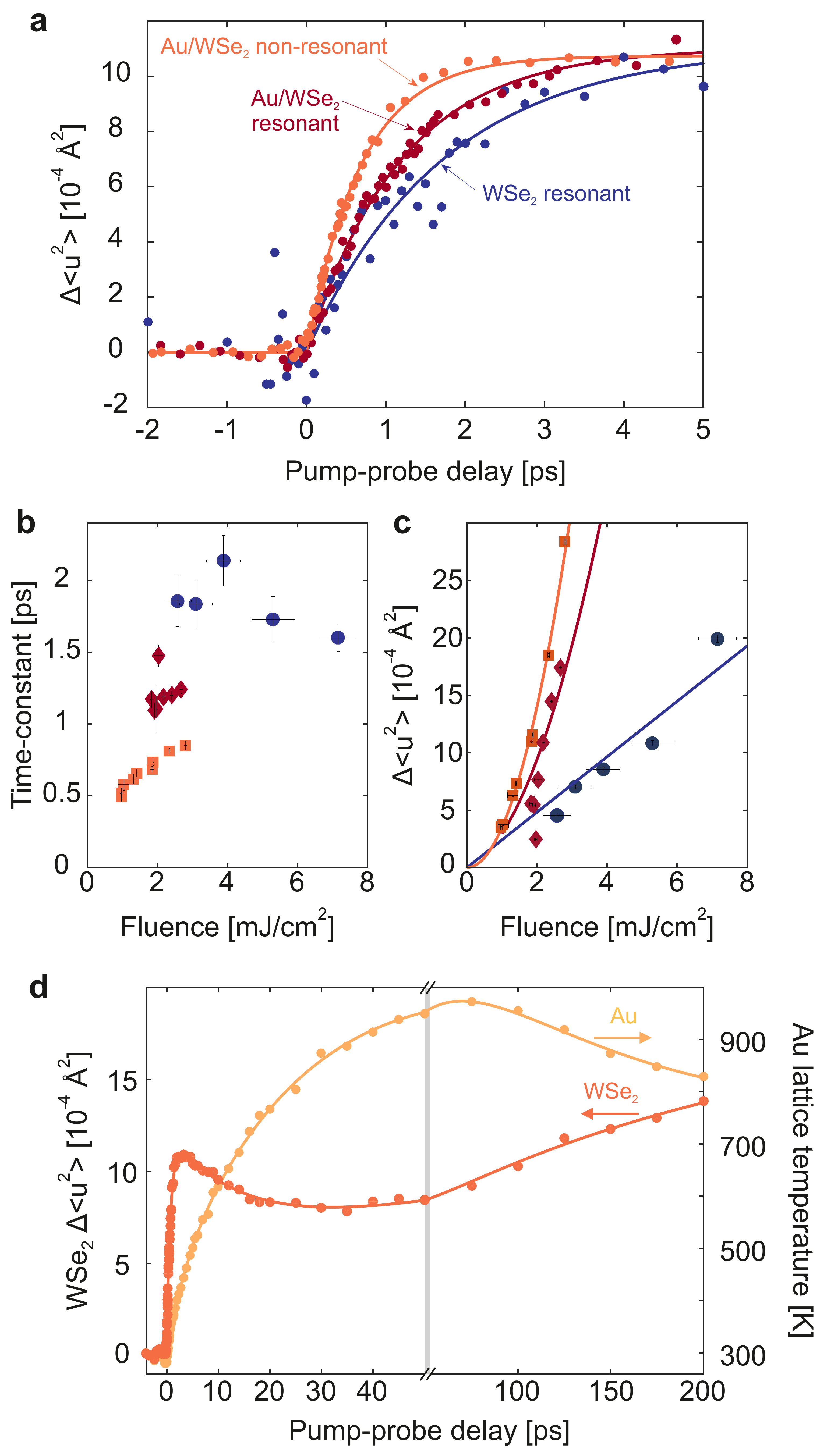}
\caption{\textbf{Ultrafast lattice dynamics of Au/WSe$_2$ heterostructures.} \textbf{(a)} Enhancement of atomic MSD due to carrier-lattice coupling as a function of the pump-probe delay for resonant excitation of pure WSe$_2$, resonant excitation of Au/WSe$_2$, and non-resonant excitation of Au/WSe$_2$ shown with blue, red, and orange points, respectively. The corresponding solid lines are fits with exponential decay functions of the form: $A\cdot exp(-t/\tau)$, where $\tau$ is the time-constant and A is the maximum atomic MSD caused by carrier-lattice relaxation. 
\textbf{(b)} and \textbf{(c)} Time-constant for carrier-lattice coupling and maximum MSD after carrier-lattice equilibration, respectively (same color codes). \textbf{(d)} Long delay dynamics of the atomic MSD of WSe$_2$ (orange curve and left axis) and the corresponding lattice temperature evolution of Au (gold curve and right axis).}
\label{fig:fig_5}
\end{figure}

Fig.~\ref{fig:fig_5}~a, shows the $\Delta\langle u^2\rangle$ of WSe$_2$ in the first 5 ps after photoexcitation for three sets of experiments: pure WSe$_2$ pumped in resonance with the A-exciton ($\lambda$=763 nm), Au-decorated WSe$_2$ ($\lambda$=763 nm), and Au-decorated WSe$_2$ with sub-band-gap ($\lambda=850$ nm) excitation \textemdash all with similar excitation density (see Methods). The time-constants for lattice heating in response to resonant excitation drops from $1.73\pm0.16$~ps in pure WSe$_2$, to $1.19\pm0.3$~ps in Au-decorated WSe$_2$,  and down to $0.68\pm0.02$~ps for sub-band-gap excitation. Pure WSe$_2$ does not show measurable lattice heating with 850 nm pumping (see SI Sect. 14). From the measurements of Fig.~\ref{fig:fig_5}~a, it becomes obvious that Au-decoration accelerates carrier-lattice equilibration in WSe$_2$ and enables absorption of sub-band-gap photons, giving rise to even faster lattice dynamics.

The measurements associated with the data presented in Fig.~\ref{fig:fig_5}~a have been repeated for various fluences. The extracted time-constants ($\tau$) and amplitudes of the fitted exponential decays are plotted as a function of the incident laser fluence in Fig.~\ref{fig:fig_5}~b and Fig.~\ref{fig:fig_5}~c, respectively. Pure WSe$_2$ has a time-constant of $\approx$1.6~ps at 7 mJ/cm$^2$ to $\approx$2 ps at 3 mJ/cm$^2$ (Fig.~\ref{fig:fig_5}~b, blue data points). With Au decoration, the WSe$_2$ lattice response to the A-exciton becomes significantly faster and all measured time-constants cluster around 1.2~ps within the probed fluence range (Fig.~\ref{fig:fig_5}~b, red data points). This is in clear agreement with the results of tr-ARPES experiments in Fig.~\ref{fig:fig_3}~c. The accelerated lattice dynamics of WSe$_2$ results from charge-injection from Au, as the hot carriers reside only for short time in the semiconductor (Fig.~\ref{fig:fig_3}~b). For pure WSe$_2$ pumped at the A-exciton resonance, the fluence dependence of the maximum MSD reached by carrier-lattice equilibration is close to linear (blue solid line, Fig.~\ref{fig:fig_5}~c). When WSe$_2$ is covered by Au, the rise of the MSD is enhanced and it assumes a non-linear fluence dependence (Fig.~\ref{fig:fig_5}~c). 

Further acceleration of carrier-lattice relaxation is observed for sub-band-gap (850 nm) excitation (Fig.~\ref{fig:fig_5}~b, orange data points). The time-constants are on the sub-picosecond timescale and as short as 500~fs at 1 mJ/cm$^2$. The non-linear fluence dependence of the maximum MSD can be represented with a power law of exponent 2.0±0.1. Thus, at the high fluences used in FED, which are all higher compared to the $\lambda=800$ nm tr-ARPES experiment, absorption of sub-band-gap photons is dominated by two-photon absorption, in line with the observations in ARPES for the $\Sigma$ valley population in Fig.~\ref{fig:fig_3}~d. 

Figure~\ref{fig:fig_5}~d shows the lattice dynamics of Au and WSe$_2$ up to 200~ps after photoexcitation with sub-band-gap light. The MSD of Au is used to extract its lattice temperature evolution through the Debye-Waller factor. Both materials show a lattice response that can be approximated with multiexponential fitting. Regarding WSe$_2$, the MSD is first rising due to carrier-lattice relaxation with $\tau_1=0.74 \pm0.02$ ps, followed by a decrease with $\tau_2=16 \pm1$ ps, which we assign to phonon-phonon equilibration in WSe$_2$~\cite{Waldecker_PRL_2017}. 

In the 50-200 ps time interval, the atomic MSD of WSe$_2$ is again rising (by 5.3$\cdot10^{-4} \AA^2$, Fig.~\ref{fig:fig_5}~d).
The lattice dynamics of Au during the first 50 ps have bi-exponential behavior with $\tau_1=4\pm1$ ps, and $\tau_2=35 \pm5$ ps. The fast, lattice-heating process is in the typical timescale (3-6 ps) for electron-phonon coupling in nano-Au, while the second is surprisingly slow for intrinsic carrier-lattice equilibration~\cite{Vasileiadis_2018,Vasileiadis_2019}. The slow heating can arise from the transfer of lattice-equilibrated dark excitons~\cite{Bertoni_2016} from WSe$_2$ towards Au and their dissociation to metallic \textit{sp}-states~\cite{Cabo_2015_MoS2onAu}. Each of these events releases $\approx$1.5~eV (the indirect band gap of WSe$_2$) and generates hundreds of vibrational quanta. The maximum lattice temperature of Au is $\approx$ 970 K, while in the 50-200 ps time-interval it cools by 140 K (Fig.~\ref{fig:fig_5}~d). The cooling of Au and the heating of WSe$_2$ at long time-delays is interpreted as re-equilibration of the two components by vibrational coupling, previously found in heterostructures of Au nanoclusters on various substrates~\cite{Vasileiadis_2018,Vasileiadis_2019}.  

\section{Discussion}

\begin{figure}[t!]
\centering\includegraphics[width=0.75\textwidth]{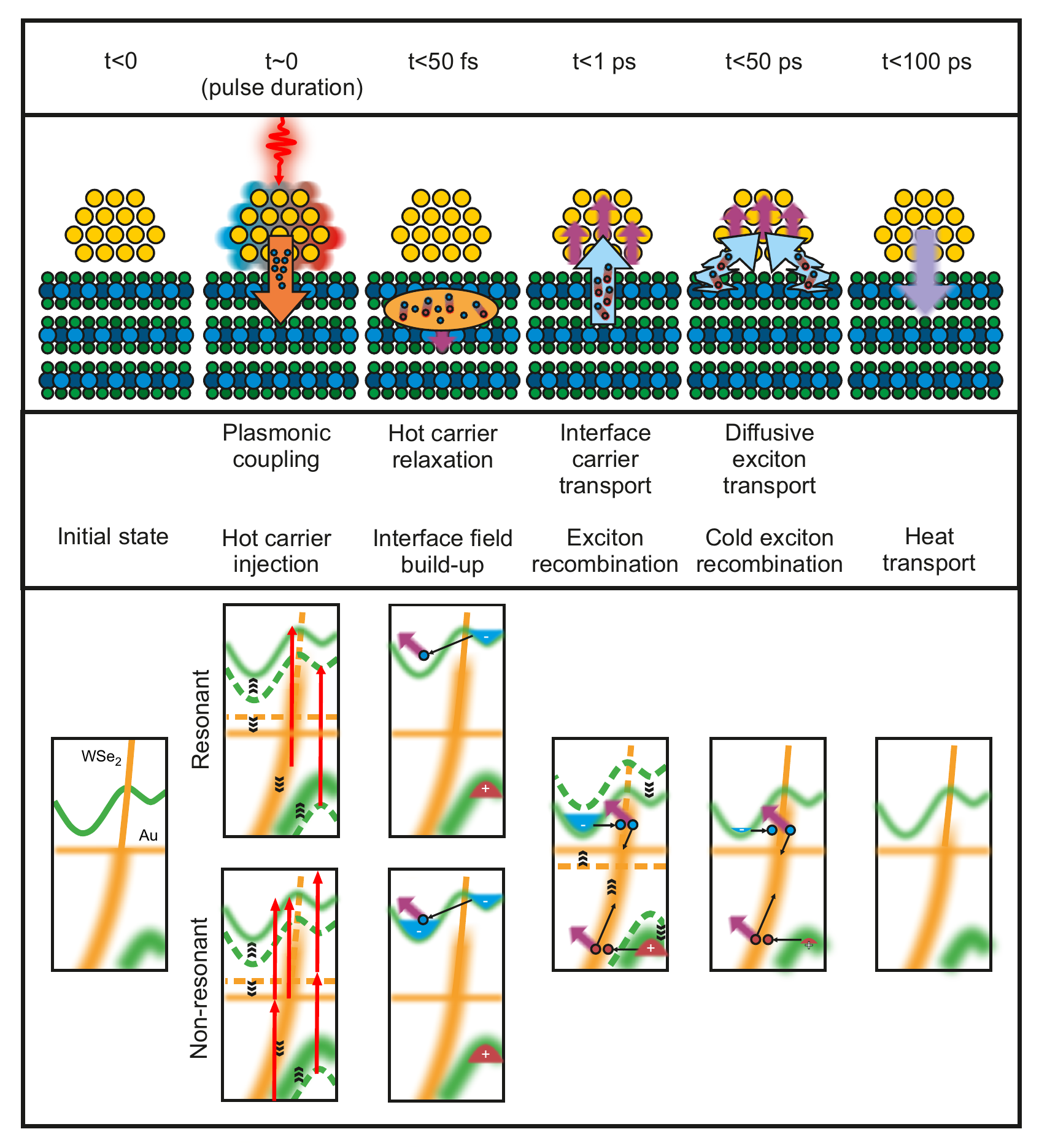}
\caption{\textbf{Energy flow across the Au/WSe$_2$ heterostructure.} Schematic sequence of events and quasiparticle transfer unfolding in Au/WSe$_2$ heterostructure after NIR excitation. Top band shows a cartoon, real space picture. Solid orange arrows represent plasmon-induced hot-carrier transfer. Solid light blue arrows indicate hot exciton transfer. Wiggly light blue arrows represent diffusive exciton recombination via metallic Au states. Fuzzy purple arrows indicate hot phonon generation by electron-phonon coupling. The fuzzy lilac arrow shows the direction of vibrational energy transfer at long timescales. The bottom panel shows the correspnding electronic structure picture, with the two different cases of resonant and sub-bandgap pumping separated at short timescales. Red arrows indicate the relevant vertical optical transitions.}
\label{fig:fig_6}
\end{figure}

Considering all experimental results, we can confirm the presence of a strong exciton-plasmon interaction, determine its microscopic  origin, and assess the kinematics of the multi-directional energy flow across the interface.
Figure ~\ref{fig:fig_6} summarizes the observed processes in the Au/WSe$_2$ heterostructure in real-space (upper panels) and momentum-space (lower panels). 
In the interpretation of the data, it is important to notice that, while FED probes the entire volume of the sample, tr-ARPES is extremely surface sensitive: for this reason we focus on observables that are homogeneous at the scale of a nanoparticle or a gap ($\approx$ 10 nm) at the timescales of interest ($>$ 10 fs), or we use the excitation wavelength to be intrinsically selective of the hot carriers while tracking population dynamics (see SI Sect. 10).

Optical illumination induces LSP at the nanoisland's surfaces and edges. In Au, plasmons produce highly energetic hot carriers that scatter across the interface on extremely short timescales ($<$10s fs). Both electrons and holes are transferred, but the asymmetric SB produces an excess electron population. In the WSe$_2$ gaps not covered by Au, the strong near-fields generate electrons and holes through non-linear multiphoton absorption: this gains relevance over hot-carrier injection or single-photon absorption at wavelengths $\geq 850$ nm and fluences $\geq 1$ mJ/cm$^2$. 

After excitation, most of the injected hot carriers lose their energy in WSe2 forming indirect (optically dark) excitons. Initially, when the carriers enter the semiconductor ballistically, they may undergo electronic scattering processes such as electron-electron, electron-hole and surface/defect scattering to relax to a thermalized hot carrier distribution.  Subsequently, the hot carriers transfer energy to the lattice predominantly via electron-phonon scattering, with possible contributions of surface/defect-assisted collisions. Singling out the individual contributions of each mechanism is beyond the scope of this work, and would require targeted experiments. We will therefore refer to these transitional phases with the general terms \textit{hot carrier relaxation} and \textit{electron-phonon relaxation}.

Then, the hot unbalanced free electrons, confined by static and dynamic interfacial fields to the vicinity ($<$ 1 nm) of the nanoisland, flow back to Au, re-equilibrating charge on a timescale of $240\pm 30$ fs. The electronic temperature of Au raises only through carrier backflow on hundreds of femtoseconds timescales, in stark contrast with the classic model of plasmonics assuming thermalization of the metallic hot carriers before interfacial transfer.

In parallel, the hot carriers generated by non-linear multiphoton absorption, distributed in the crystal, turn into dark excitons by electron-phonon coupling with the lattice at sub-picosecond timescales due to their large scattering phase-space. Both of these processes shorten the electron-lattice relaxation time of WSe$_2$ down to $500$ fs for low fluence ($\leq 1$ mJ/cm$^2$) and sub-bandgap ($\geq 850$ nm) pumping. After electron-lattice relaxation in WSe$_2$ (t$>$1 ps), the remaining cold dark excitons move diffusively until they reach the nanoislands, where they dissociate and produce a second intense lattice-heating of Au on a timescale of $35\pm5$ ps. This process is maximized when using sub-bandgap pumping at high fluences, where nonlinear multiphoton effects generate large populations of dark excitons in the plasmonic hotspots appearing in a few tens of nanometers radius from the interface.

In conclusion, we have disentangled the contributions of individual materials and specific plasmonic, electronic and phononic excitations to interfacial energy transfer processes. We have demonstrated that strong plasmon-exciton interaction leads to immediate energy transfer to the semiconductor, with the nanometal being heated by carrier backflow and exciton recombination at two distinct timescales. The two-stepped energy backflow arises from the presence of two species of excitons: the ones generated by strong plasmon-exciton interaction in the vicinity of the nanoisland and the ones produced by non-linear absorption in the plasmonic hot-spots. 

The Au/WSe$_2$ system is close to real-life applications, with the Au nanoislands offering plasmonic and catalytic properties and the WSe$_2$ displaying a rich excitonic structure that allows following different energy paths at different photon energies. The possibility of harvesting plasmonic energy into high energy excitons is a mechanism useful for optoelectronics and photochemistry ~\cite{Manjavacas_2014}, and the backflow arising from recombination might be minimzed by careful engineering of the band alignment and momentum-matching conditions. The subsequent energy flow causing intense, local, lattice heating of nanoscale Au is a quasi-thermal process that can be used for photothermal conversion and for catalyzing chemical reactions~\cite{Thermal_effects_in_Photocatalysis}. These findings offer new possibilities to tune the quasi-thermal response by controlling the exciton population, possibly with means other than multi-wavelength and fluence excitation protocol employed here. For example, interfacing Au nanoislands with more complex 2D semiconducting heterostructures~\cite{Jin_2018} might allow electrical control of the exciton population and thus of photochemical performance. Even further, manipulating the valley degree of freedom might give access to band topology-controlled functionalities~\cite{Li_2020, Lin2022}.

\section{Experimental Section}

\threesubsection{Time-resolved ARPES}
The time-resolved ARPES experiments were performed using a home-built optical parametric chirped-pulse amplifier (OPCPA) with 500 kHz repetition rate~\cite{Puppin_2015}. The OPCPA is used to drive high-order harmonic generation (HHG) by tightly focusing laser pulses onto a Argon gas jet. The HHG produces a comb of odd harmonics of the driving laser, extending up to the 11th order. The co-propagating fundamental is separated from the XUV harmonic beam using reflection onto a silicon wafer at Brewster's angle. A single harmonic (7th order, 21.7 eV, p-polarized, pulse duration: 23$\pm$4 fs FWHM, energy width: 110 meV FWHM) is isolated by reflection off a focusing multilayer XUV mirror and transmission through a 400 nm thick Sn metallic filter. A photon flux of up to 2x10$^{11}$ photons/s at the sample position is achieved~\cite{Puppin_2019}. The 800 nm pump was s-polarized, had pulse duration of 36$\pm$4 fs, incident fluence of 750 $\pm 50 \mathrm{\mu J}$. The 1030 nm pump was s-polarized, had pulse duration of approx. 250 fs, and variable incident fluence. The bulk $\mathrm{WSe_2}$ samples are handled by a 6-axis manipulator (SPECS GmbH). The data are acquired using a time-of-flight momentum microscope (METIS1000, SPECS GmbH) and processed using custom-built code~\cite{Xian_2020} for Fig.~\ref{fig:fig_2} c, while all the dynamics was measured using a hemispherical electron spectrometer (PHOIBOS150, SPECS GmbH) to achieve higher statistical performance~\cite{Maklar_2020}.

\threesubsection{Femtosecond Electron Diffraction}
The FED apparatus employs electron pulses to measure the ultrafast lattice dynamics in response to photoexcitation. The diameters of the probed and pumped (photoexcited) areas are 100 $\mu$m and 400 $\mu$m, respectively. For each diffraction peak, the intensity is extracted by direct integration of the total counts within a circle of 20 pixel diameter. The center of the circle coincides with the center of mass of the peak and it is recalculated for each diffraction pattern in order to eliminate the effect of instabilities of the electron gun and the magnetic lens. An ultrashort laser pulse (100 fs) of selected wavelength (TOPAS Prime NirUVis) excites the electronic subsystem. The lattice response is probed with a time-resolution in the order of 300 fs using ultrashort, high energy (60 keV) electrons pulses~\cite{Waldecker_2015} that impinge on the thin, freestanding sample, producing a diffraction pattern in transmission (Fig.S9a). After the arrival of an ultrashort laser pulse, the intensity of all diffraction peaks ($I_{hkl}$) decreases, and the inelastic scattering background increases, due to the generation of phonons by excited charge carriers and the Debye-Waller effect.

The relative decay of the diffraction peaks is used to calculate the time-dependent change of the atomic MSD ($\Delta \langle u^2\rangle$) through the formula:
\begin{eqnarray}
   \Delta\langle u^2\rangle =-\frac{3d_{hkl}^2}{4\pi^2}\ln{\dfrac{I_{hkl}(t)}{I_{hkl}(t<0)}}
\label{eq:one},
\end{eqnarray}
where $d_{hkl}$ is the spacing between crystal planes for each diffraction peak. For each measurement the $\Delta \langle u^2\rangle $ is averaged over all the diffraction peaks. Thus, the present analysis does not explicitly take into account nonthermal lattice modes~\cite{Waldecker_NLM} and the distinct vibrational amplitudes for different types of atoms in compounds, since the aim is to compare the average MSD dynamics~\cite{Waldecker_PRL_2017} for pure and Au-decorated WSe$_2$. For this purpose, the measurements of Fig.\ref{fig:fig_5}~a have been performed adjusting the fluence in order to obtain the same maximum atomic MSD of $(10.99 \pm0.09)\cdot10^{-4} \AA^2$. For pure WSe$_2$ the incident laser fluence is (5.3 ±0.6) mJ/cm$^2$, while for Au-decorated WSe$_2$ it is $(2.18 \pm0.04)$ mJ/cm$^2$ for A-exciton pumping and $(1.88 \pm0.02)$ mJ/cm$^2$ for sub-band-gap photoexcitation.

\threesubsection{Sample preparation}
Samples for ARPES were prepared by cleaving bulk WSe$_2$ crystals in vacuum (base pressure better than 5x10$^{-11}$ mbar). The crystals were then cooled to 70 K, then Au was evaporated on the surface for 5 min at a calibrated rate of $2\pm1\AA$/min. The deposition at low temperature ensures homogeneous coverage across the sample. The sample was then "annealed" at 300 K for 30 min to enable island formation before measurement. The average height of the nanoislands is calculated at $2\pm1$ nm by considering nominal film thickness ($1\pm0.5$ nm) and average area coverage ($\approx 50\%$). 

The multilayer free-standing membranes for the FED measurements were prepared by exfoliation from bulk single crystals (HQ Graphene). Large flakes of WSe$_2$ are first separated from the bulk single crystal with a lancet. Then the flakes are attached on a glass substrate with a water soluble glue (Crystalbond) and thinned down by exfoliation with a scotch tape. The thin areas of WSe$_2$  are then separated from the substrate with a scalpel and placed into water to separate WSe$_2$  from the glue. Finally, the floating flakes of WSe$_2$ are scooped out with a TEM copper grid held by a tweezer and left to dry. For the Au/WSe$_2$, the TEM copper grid with the flake has been placed in a UHV chamber and 2 nm of Au were evaporated on top with electron-beam-evaporation using a rotating sample holder for homogeneous deposition. 

\threesubsection{Optical measurements}
Using optical microscopy we selected a spot on a thin flake of bare WSe$_2$ ($\sim$20 nm) suspended over a TEM grid, choosing a flat (without wrinkles) part of the sample. For this point of the sample we have recorded the optical transmission and reflectance spectra ($T$ and $R$, respectively) using a micro-absorbance spectrometer, a supercontinuum laser (FIANIUM) as the broadband light source, and a fiber spectrometer (Avantes). In all cases we have recorded dark and reference spectra. The absorption spectrum ($A$ in \%) was calculated as A=100-T-R. Subsequently, the flake has been decorated with 2 nm thick Au and the measurements have been repeated at the same spot on the sample. 

\threesubsection{Finite element calculations}
To understand the distribution of fields at the surface and explore the linear response of the heterostructure, we simulated the optical response in the frequency domain using the Optics package in the commercial finite difference software COMSOL Multiphysics.
A block of 100 nm x 200 nm x 500 nm was used to model the substrate, while 2 nm thick islands were extruded on the surface using a lateral profile extracted from a micrograph, to exactly match the effective spatial distribution. A 100 nm x 200 nm x 500 nm vacuum layer was added on top. The optical field was incident vertically with polarization along the x direction (maps in Fig.~\ref{fig:fig_1} are rotated 90°). The mesh was optimized to achieve minimum element quality (skeweness) $>$0.1. The cuboid is surrounded by periodic boundary conditions, except for the input and output planes (along the z direction), where perfectly matched layer conditions were used. 

To evaluate the LSP-generated hot carrier photocurrent we integrate the expression:
\begin{align}
    \label{eq:eq1}
    P = \int_{V_{Au}} |E|^2 d\mathbf{r}\ ,
\end{align}
over the nanoparticle volume. As discussed in~\cite{Zheng_2015}, this is proportional to the number of LSP-generated hot-carriers crossing the interface in a Schottky junction. To calculate the local the field enhancement $|E|/|E0|$, we have simply evaluated the ratio between $|E|$, the modulus of the electric field in the heterostructure and $|E0|$ the modulus of the electrical field in absence of Au.

The presence of both hot-carrier dominated photoabsorption and field enhancement is robust against variations in the shape and thickness of the islands, including islands with Winterbottom shape that represents the equilibrium configuration at high temperatures~\cite{Reidy2022}, as long as the average coverage of 50\% with typical sizes of islands and gaps around 10 nm are preserved.

\threesubsection{Statistical analysis}
Time-of-flight generated multidimensional ARPES data have been preprocessed by binning the tabular data structure into 4-dimensional hypervolumetric data. During the binning procedure, image distortion correction, image registration, momentum, energy and pump-probe delay calibrations are applied according to the procedure described elsewhere~\cite{Xian_2020}.
Hemispherical analyzer ARPES data do not require binning. Calibration-based distortion correction, together with momentum, energy and pump-probe delay calibration have been applied in the preprocessing phase. The pump-probe time-traces were averaged over multiple scans of the delay stage. Data are presented using mean ± SD, i.e. 68\% confidence intervals. The fitting procedures employed are based on the non-linear least square method. Data analysis is performed using custom built routines in the proprietary software IgorPro.

The raw diffraction patterns were corrected by dark image subtraction (electron beam off and same exposure time) and flat field correction (obtained with homogeneous illumination of the electron camera with a strongly defocused polycrystalline diffraction pattern). The intensity of each Bragg spot was integrated over a circular area, whose center matched the position of local maximum intensity. The pump-probe time-traces were averaged over multiple scans of the delay stage. The time-constants and amplitudes, extracted by nonlinear least squares fitting of exponential decay functions, have error bars representing the 68\% confidence intervals. Data analysis is performed with custom built Matlab scripts.

\medskip
\textbf{Supporting Information} \par 
Supporting Information is available from the Wiley Online Library or from the author.

\medskip
\textbf{Acknowledgements} \par 
This work was funded by the Max Planck Society, the European Research Council (ERC) under the European Union's Horizon 2020 research and innovation and the H2020-EU.1.2.1. FET Open programs (Grant Numbers: ERC-2015-CoG-682843, ERC-2015-AdG-694097, and OPTOlogic 899794), the Max Planck Society's Research Network BiGmax on Big-Data-Driven Materials-Science, and the German Research Foundation (DFG) within the Emmy Noether program (Grant Number: RE 3977/1), through SFB 951 “Hybrid Inorganic/Organic Systems for Opto-Electronics (HIOS)” (Project Number: 182087777, projects B12 and B17), the SFB/TRR 227 “Ultrafast Spin Dynamics” (projects A09 and B07), the Research Unit FOR 1700 “Atomic Wires” (project E5), and the Priority Program SPP 2244 (project 443366970). Tommaso Pincelli acknowledges financial support from the Alexander von Humboldt Foundation. Thomas Vasileiadis acknowledges support from the Marie Skłodowska-Curie widening fellowship (101003436 - PLASMMONS). Thomas Vasileiadis and Emerson Coy thank Prof. Stefan Jurga (NanoBioMedical Centre, AMU Poznań) for the use of the HR-TEM instrument. Emerson Coy Acknowledge the partial financial support from the National Science Centre (NCN) of Poland by the OPUS grant 2019/35/B/ST5/00248. Samuel Beaulieu acknowledges financial support from the NSERC-Banting Postdoctoral Fellowships Program. Niclas S. Mueller acknowledges support from the German National Academy of Sciences Leopoldina through the Leopoldina Postdoc Scholarship.

\medskip

%
\bibliographystyle{MSP}
\bibliography{Main}

\end{document}



\pagestyle{fancy}
\rhead{\includegraphics[width=2.5cm]{vch-logo.png}}

\title{Observation of multi-directional energy transfer\\
in a hybrid plasmonic-excitonic nanostructure\\
Supplementary Information}

\maketitle


\author{Tommaso Pincelli* $^\dagger$}
\author{Thomas Vasileiadis $^\dagger$}
\author{Shuo Dong}
\author{Samuel Beaulieu}
\author{Maciej Dendzik}
\author{Daniela Zahn}
\author{Sang-Eun Lee}
\author{Hélène Seiler}
\author{Yinpeng Qi}
\author{R.Patrick Xian}
\author{Julian Maklar}
\author{Emerson Coy}
\author{Niclas S. Mueller}
\author{Yu Okamura}
\author{Stephanie Reich}
\author{Martin Wolf}
\author{Laurenz Rettig}
\author{Ralph Ernstorfer*}\\
$^\dagger$ These authors contributed equally.


\dedication{}

\begin{affiliations}
Dr. T. Pincelli, Dr. T. Vasileiadis, Dr. S. Dong, Dr. S. Beaulieu, Dr. M. Dendzik, Dr. D. Zahn, S.-E. Lee, Prof. H. Seiler, Dr. Y. Qi, Dr. R. P. Xian, J. Maklar, Prof. M. Wolf, Dr. L. Rettig, Prof. Ernstorfer\\
Fritz-Haber-Institut der Max-Planck-Gesellschaft, Faradayweg 4-6, 14195 Berlin, Germany\\
Email Address: pincelli@fhi-berlin.mpg.de, ernstorfer@tu-berlin.de\\

Prof. H. Seiler, Dr. N. S. Mueller, Y. Okamura, Prof. S. Reich\\
Freie Universität Berlin, Arnimallee 14, 14195 Berlin, Germany.\\

Dr. T. Pincelli, Prof. R. Ernstorfer\\
Institut für Optik und Atomare Physik, Technische Universität Berlin, Straße des 17.~Juni 135, 10623 Berlin, Germany\\

Dr. T. Vasileiadis\\
Faculty of Physics, Adam Mickiewicz University, Uniwersytetu Poznanskiego 2, 61-614 Poznan, Poland\\

Dr. S. Beaulieu\\
Université de Bordeaux - CNRS - CEA, CELIA, UMR5107, F33405, Talence, France.\\

Dr. M. Dendzik\\
Department of Applied Physics, KTH Royal Institute of Technology, Hannes Alfvéns väg 12, 114 19 Stockholm, Sweden.\\

Dr. Y. Qi\\
Center for Ultrafast Science and Technology, School of Physics and Astronomy, Shanghai Jiao Tong University, 200240 Shanghai, China.\\

Dr. R. P. Xian\\
Department of Statistical Sciences, University of Toronto, 700 University Avenue, Toronto, M5G 1Z5, Canada.\\

Dr. E. Coy\\
NanoBioMedical Centre, Adam Mickiewicz University, ul. Wszechnicy Piastowskiej 3, PL 61614 Poznań, Poland.\\

Dr. N. S. Mueller\\
NanoPhotonics Centre, Cavendish Laboratory, Department of Physics, University of Cambridge, JJ Thomson Avenue, Cambridge CB30HE, United Kingdom.\\
\end{affiliations}


\keywords{hybrid plasmonics, time resolved ARPES, femtosecond electron diffraction, interfacial
charge transfer, 2D semiconductors.}



\justifying
\section{Transmission electron microscopy and particle shape analysis}

High resolution Transmission electron studies were performed in an aberration corrected JEOL - ARM200F, working at an accelerating voltage of 200 kV. The samples were mounted in a reinforced beryllium holder (JEOL) and left in dark conditions and vacuum over night. Images were collected with a zone axis WSe$_2$ [0,0,1]. 

The following analysis shows that the Au covered area fraction is 50\%, and that in-plane particle size follows a skewed distribution whose median is $10$ nm. The nominal thickness of the nanoparticles is estimated to be around 2 nm. 
The epitaxial relationship is evident from the static diffractogram reported in Fig.~1d of the main text, where single crystalline Bragg peaks from Au are observed as a replica for each WSe$_2$ substrate peak. The slight offset between the peaks results from the mismatch between the two lattice structures that, combined with the large inelastic mean free path of Au atoms on WSe$_2$ surface, is at the origin of the Vollmer-Weber growth resulting in the self-assembled nanostructures~\cite{Rettenberger_1998}. 

\begin{figure}[h!]
\centering\includegraphics[width=0.75\textwidth]{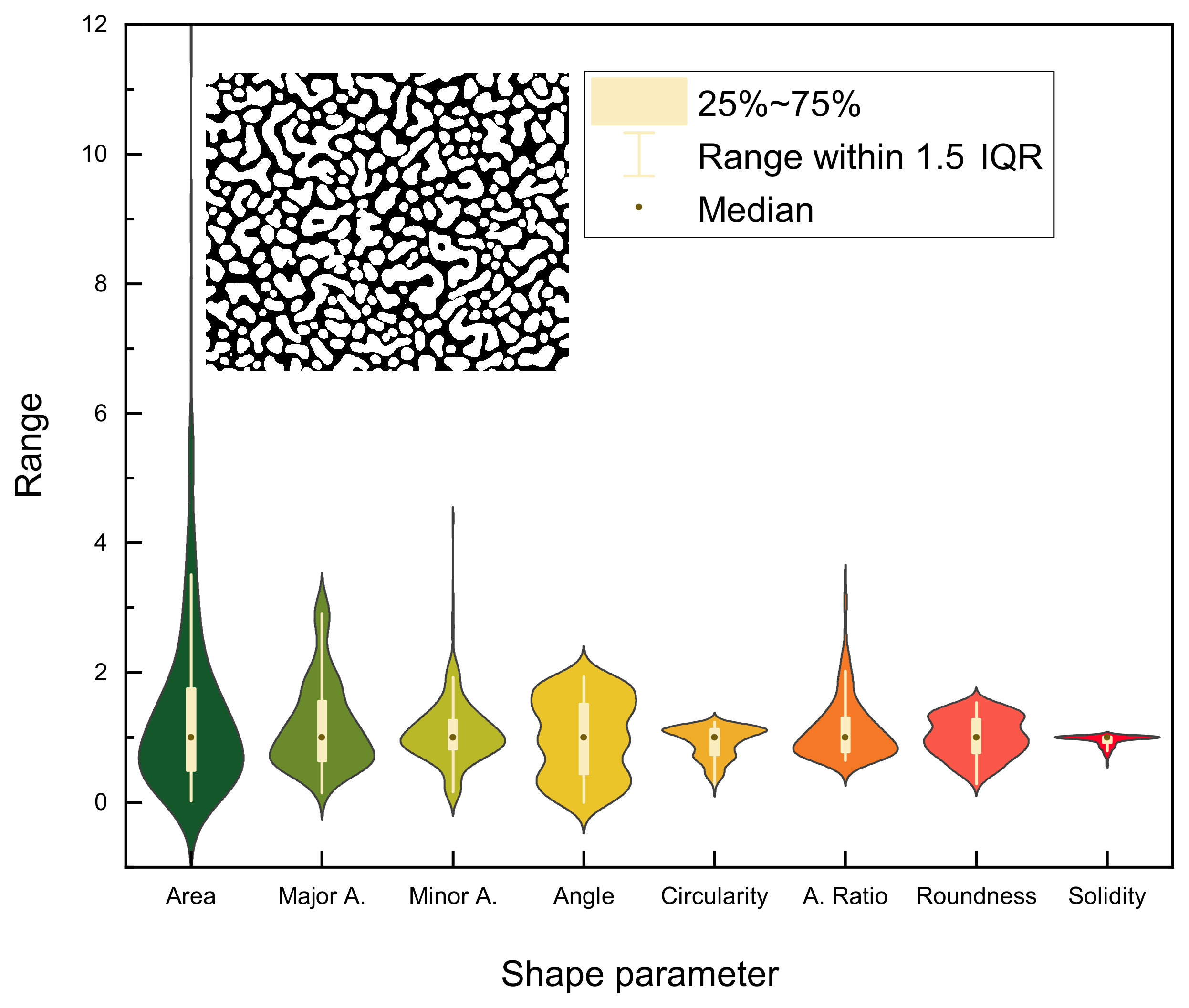}
\caption{Violin plot showing the distributions of the values of each parameter. The vertical scale of each distribution has been normalized to the mean value, reported in Tab.~\ref{tab:tab_S1}. The "range" axis thus represents the extent of the distribution in units of the mean value. The thick line encloses the range between 25\% and 75\% of the mean value, while the thin line marks the range within 1.5 times the interquartile range.}
\label{fig:fig_S1}
\end{figure}

The  particle shape analysis was performed using a TEM microscopy image from the same sample and instrument as in Fig.~1d of the main article. The image spans a range of 180 nm x 213 nm. The image has been thresholded to define the islands. The result is reported in the inset in Fig.~\ref{fig:fig_S1}. The Au islands are in white. The area and circularity of the nanoparticles were extracted using the ImageJ automated analysis software. Circularity is given by $4\pi*Area/(Perimeter)^2$. The nanoparticles were also fitted with ellipses, allowing to extract several parameters: major and minor axes, the angle of the major axis with respect to the horizontal (Angle), and the aspect ratio.
The results are summarized in Tab.~\ref{tab:tab_S1} and Fig.~\ref{fig:fig_S1}.

\begin{table}
\begin{tabular}[h!]{@{}llllllllll@{}} 
\hline
Data & Mean & Standard Deviation & Variance & Skewness & Kurtosis & Median \\
\hline
Area & 89.67 $\mathrm{nm^2}$ & 93.93 $\mathrm{nm^2}$ & 8823.18 $\mathrm{nm^4}$& 2.92 & 12.62 & 63.08 $\mathrm{nm^2}$\\
Major & 12.86 $\mathrm{nm}$ & 7.38 $\mathrm{nm^2}$ & 54.52 $\mathrm{nm^2}$ & 1.07 & 0.61 & 10.87 $\mathrm{nm}$\\
Minor & 7.39 $\mathrm{nm}$ & 3.43 $\mathrm{nm}$ & 11.73$\mathrm{nm^2}$ & 1.72 & 8.39 & 6.88 $\mathrm{nm}$\\
Angle & 92.90° & 55.63° & 3094.61°$^2$ & -0.09& -1.29 & 92.72°\\
Circ. & 0.72 & 0.19 & 0.04 & -0.84 & -0.31 & 0.78\\
AR & 1.79 & 0.77 & 0.59 & 1.91 & 4.62 & 1.58\\
\hline
\end{tabular}
\caption{Table reporting the values of the statistical analysis of particle shape. The total number of particles examined is 261, as observed in the overview image inset in \ref{fig:fig_S1}. In each row, all the statistical parameters are reported for: nanoparticle area, the major and minor axis of the best fitting ellipse, the angle from horizontal of the best fitting ellipse major axis, the degree of circularity and the aspect ratio.}
\label{tab:tab_S1}
\end{table}

From Tab.~\ref{tab:tab_S1}, emerges a simplified average shape of thin, ellipsoidal disks with major axis around 13 nm and minor axis around 7 nm, elongated in randomly oriented directions. By looking at the distributions of Fig.~\ref{fig:fig_S1}, however, it is clear that the sizes and shapes of the nanostructures span a vast range of scales, thus making necessary a true-to-shape finite-element modelling, as reported in Fig.~\ref{fig:fig_S4}. 

\section{Localized fields and finite elements calculations}

We report the distribution of the field enhancement $|E|/|E_0|$ (where $|E|$ is the modulus of the electric field in the heterostructure and $|E_0|$ is the field in absence of Au). This scalar quantity is calculated by performing finite element, frequency domain calculations on a model Au/WSe$_2$ heterostructure (see also Methods in the main article). The dielectric functions were from tabulated data: Au from \cite{Johnson_1972} and WSe$_2$ from \cite{Gu_2019}. To calculate $|E_0|$, the calculation was repeated with identical settings except for the refractive and absorption index of Au, that were set to the ones of vacuum instead. The total modulus was calculated for both $E$ and $E_0$ 3D vector fields, and their ratio determined to produce the three-dimensional scalar field of field enhancement. The field is then sampled by slicing the model volume with a plane parallel to the Au/WSe$_2$ interface. In the following images, we report the field enhancement on two planes above and below the interface at various different wavelengths.

\begin{figure}[h!]
\centering\includegraphics[width=\textwidth]{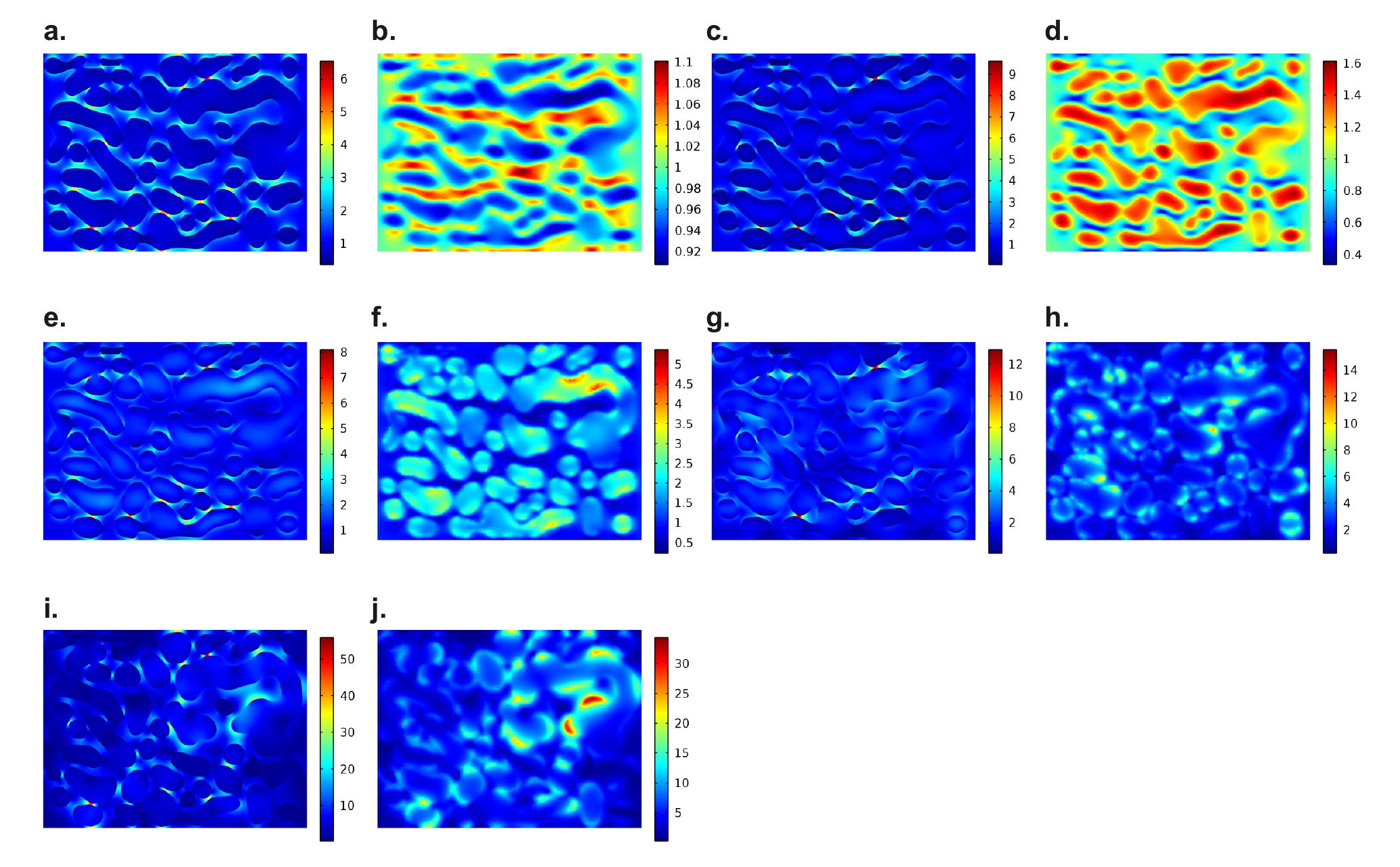}

\caption{\textbf{Electric field enhancement at various wavelengths.} a.-b. Electric field enhancement in Au nanoislands on WSe$_2$ 2nm above (a.) and 1 nm below (b.) the interface, with 400 nm excitation. c.-d. Same, but for 800 nm excitation. e.-f. Same, but for 860 nm excitation. g.-h. Same but for 1060 nm excitation. i.-j. Same, but for 1200 nm excitation.}
\label{fig:fig_S4}
\end{figure}

The slices above the surface (Fig.~\ref{fig:fig_S4}~a,c,e,g,i) show the field enhancement within the nanoislands and in the vacuum space between them. At 400 nm the excitation frequency is above the Au interband transtion, Au absorption becomes significant, and the field enhancements are largely confined outside of the nanoparticles. With longer wavelengths, we observe significant field enhancement also within the nanoparticles as the bulk shape of the particle becomes more relevant in determining the dominant plasmonic modes. In the 800-1030 nm range explored in the experiments, the generation of hot electrons is therefore enhanced by localized plasmonic excitations that affect a large fraction of the volume of the particle and depend strongly on the particle shape.  
The slices below the surface (Fig.~\ref{fig:fig_S4}~b,d,f,h,j), on the other hand, indicate how the field tailoring produced by the nanostructure propagates within the WSe$_2$ substrate. Above the semiconducting direct bandgap, the field enhancements within WSe$_2$ are small. At 400 nm light is strongly absorbed also by the nanoparticles, and we observe only weak enhancement in the gaps between them (Fig.~\ref{fig:fig_S4}~b). At 800 nm, the particles act as field concentrators and a weak enhancement is created under their footprint (Fig.~\ref{fig:fig_S4}~d).
At longer wavelengths the enhanced field can propagate more freely in the semiconductor, and we observe large field enhancements generated by the edges of the particles whose shape resonate with the excitation wavelength (Fig.~\ref{fig:fig_S4}~f,h,i). This therefore suggests that, as we move towards longer wavelengths, we increase the probability for multiphoton processes in the semiconductor due to large field enhancements.

\section{Band structure calculations with density functional theory}

For the overlayed curves in Fig.~2b (main article) and Fig.~\ref{fig:fig_S2}~c, we performed density functional theory (DFT) calculation of WSe$_2$ and Au with the projector augmented wave code GPAW~\cite{Mortensen_2005} using GLLBSE xc-functional, separately. 
The GLLBSC is an orbital-dependent exact exchange-based functional that is well suited for the description of noble metals~\cite{Yan2011,Lin2014}. While DFT+U might provide a better description of the d-states of Au~\cite{Avakyan2020}, we opted for a parameter-free functional that still provides a good, general description of Au(111) bandstructure, as the d-bands are not directly involved in the dynamics discussed in this work.
For Au, we performed a slab calculation, with 5 ML Au(111) (2.5 nm) thickness and 15 Å vacuum thickness. 
The choice of a slab calculation allowed us to obtain an ab-initio description of the Shockley surface state.
We used a Monkhorst-Pack sampling of the Brillouin zone with (12x12x1) points. The plane wave energy cutoff was at 600 eV, the occupation defined by a Fermi-Dirac distribution with 0.01 eV width.
For WSe$_2$, we performed a bulk calculation with (12x12x12) Monkhorst-Pack sampling, plane wave energy cutoff at 600 eV and occupation defined by a Fermi-Dirac distribution with 0.01 eV width. The bandpath was selected to cut the Brillouin zone at k$_z$=0.
All DFT calculations are performed with the projector augmented wave code GPAW using fully realtivistic - thus including spin-orbit coupling - plane wave basis sets and PAW potentials version 0.9.2.

\section{Weak hybridization of the electronic states}

Besides the general agreement to the DFT calculations for the two separate materials reported in Fig.2~b and Fig.~\ref{fig:fig_S2}~c, we have attempted to further investigate the effects of interfacing on the electronic structure of the two materials.

The first evidence is the absence of any variation in the shape of the core-level peaks shown in Fig.~\ref{fig:fig_S2}~a,b, which would display the evolution of substructures arising from different valence states of the atoms in the presence of alloying, chemical bonding or oxidation. The peaks of Selenium (not shown), were also measured and do not show any significant lineshape change. 

Furthermore, it is possible to inspect the valence band with a more critical insight. Two features have been identified in \cite{Bruix_2016}, where monolayers of MoS$_2$ on bulk Au(111) were studied, that showed a deviation between the suspended theoretical bandstructure of ML MoS$_2$ and the one observed on the Au(111) substrate.

\begin{figure}[h!]
\centering\includegraphics[width=\textwidth]{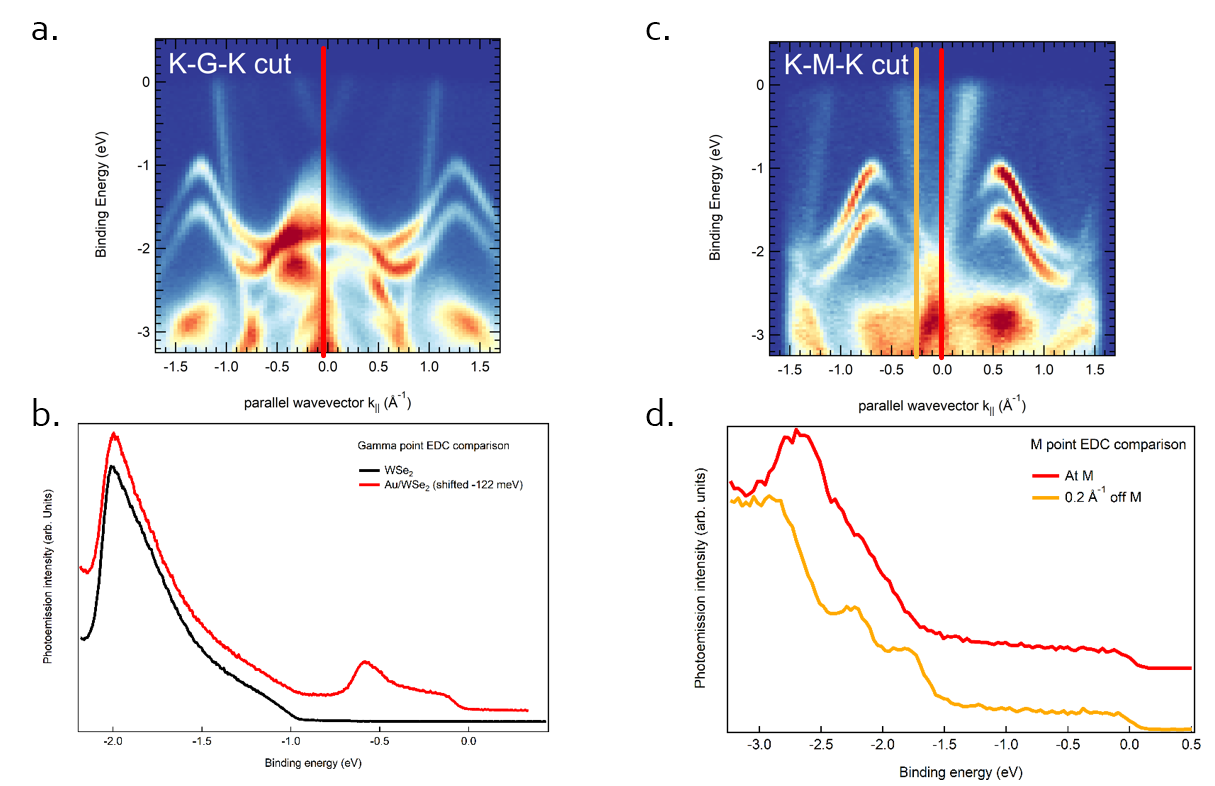}
\caption{\textbf{Weak interaction modifications of the electronic structure.} a. Static bandstructure cut along the K-$\Gamma$-K direction. Red line shows the EDC cut shown in  panel b. b. EDC cuts at the $\Gamma$ point for bare WSe$_2$ (black) and Au/WSe$_2$ (red). The red EDC has been shifted by 122 meV to compensate the effects of band alignment. c. Static bandstructure along the K-M-K direction. The yellow and red lines show the respective EDC cuts shown in panel d. d. EDC cuts at M (red) and slightly off M (yellow), showing how, moving toward the M point, the double peaked structure around 2 eV becomes a spin-degenerate feature around 2.5 eV.}
\label{fig:fig_S3}
\end{figure}

The first is a flattening and shift towards lower binding energies of the band at the $\Gamma$ point. This phenomenon arises because the orbital character in this region is formed predominantly by the chalcogenide p$_z$ orbitals and transition metal d$_{z^2}$ and $d_{yz}$ orbitals, and is therefore sensitive to out-of-plane chemical interactions. In \cite{Bruix_2016}, the MoS$_2$ valence band top at $\Gamma$ is found to be shifted 310 meV to higher binding energies due to hybridization with Au d-band continuum. No such effect is observed in the present case of thin Au(111) overlayers on WSe$_2$, as demonstrated by Fig.~\ref{fig:fig_S3}~a,b.

Another region where the effects of interfacing are seen is the M point. Indeed this represents a time-reversal invariant momentum point, where the combination of crystal symmetry and time-reversal symmetry enforce spin degeneracy. The lowest binding energy band, that is spin-orbit split at the K points, is therefore degenerate at the M point. In the case of ML MoS$_2$ on Au, such degeneracy is not observed, owing to the suppression of momentum-matching constraints at the edge of the Brillouin zone caused by strong hybridization with the Au states. In the present case, the band at the M point appears to be spin degenerate as shown in Fig.~\ref{fig:fig_S3}~c,d.

\section{Core-level photoemission and band alignment}

\begin{figure}[h!]
\centering\includegraphics[width=\textwidth]{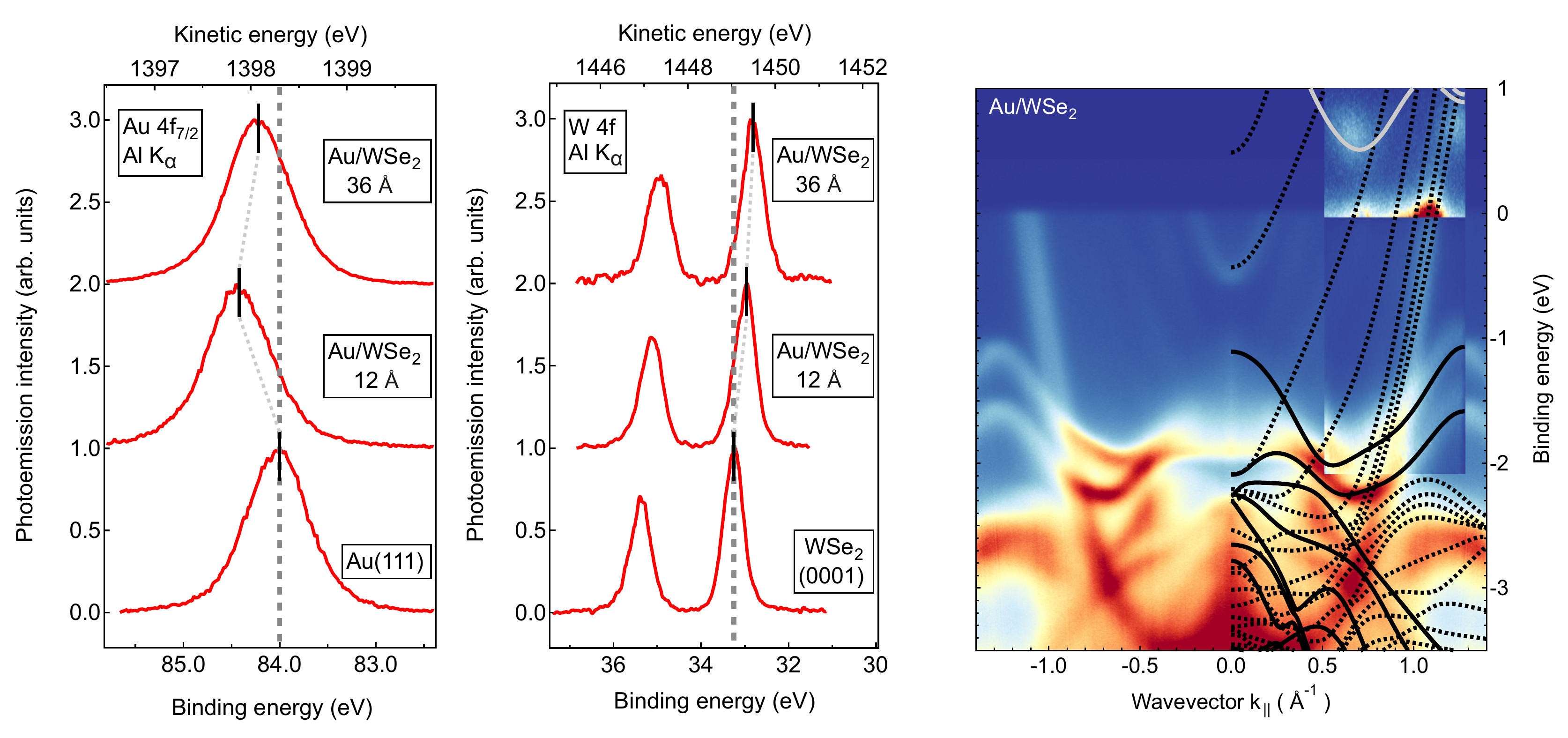}
\caption{\textbf{Band alignment.} a. Au 4f$_{7/2}$ peak at room temperature for bulk Au(111), for 12 Å Au on WSe$_2$ and for 36 Å Au on WSe$_2$. Dashed line marks the reference for Bulk. Black markers indicate peak position as obtained by fitting with Doniach-Sunjic lineshape convolved with a Gaussian. b. W 4f doublet at room temperature for freshly cleaved WSe$_2$(0001), and the heterostructure at two coverages reported in panel a. Dashed vertical line marks the reference for the bulk, black markers indicate the peak position as obtained by fitting Voigt lineshapes. c. Combined image of static ARPES, a time-resolved ARPES snapshot at t=20 fs (inset), and bandstructure calculations. note that the tr-ARPES snapshot has been aligned energetically considering the fact that the Fermi edge is shifted by -40 meV in the transient signal. }
\label{fig:fig_S2}
\end{figure}

The band alignment between Au and WSe$_2$ can be explored with very precise insight using photoemission data. Firstly, we consider the shifting of the core levels. We measure the Au 4f and W 4f peaks for the separate surfaces of Au(111) and WSe$_2$(0001), and for the heterostructure. To gain a deeper insight in the effects of Au coverage, we also measured at two different nominal Au thicknesses, 12 Å and 36 Å. The results are reported in Fig.~\ref{fig:fig_S2}.

Considering the electron affinity X$_{WSe_2}$=4.1 eV, the Fermi edge position $\Phi$=4.4 eV, and the indirect bandgap E$_g$=1.47 eV, in a freshly cleaved WSe$_2$ (0001) surface the virtual Fermi edge position is just 300 meV below the conduction band minimum, in line with the n-type behaviour observed for intrinsic WSe$_2$.
In this condition, contacting the Au(111) facet, of work function $\Phi_{Au}$=5.5 eV, would cause a shift of the Fermi level in WSe$_2$ of $\Delta\Phi$=1.1 eV. This would lead to the Fermi level being very close to the valence band, just 100-70 meV above.
However, this is not what is observed. Indeed, we only observe a shift of 170$\pm$5 meV of the WSe$_2$ bands when Au is evaporated on the surface. This indicates that the band bending is strongly suppressed and the Fermi level is moved to only 0.47 eV below the conduction band.

Considering the Schottky-Mott theory of contact potential, the Fermi level would be expected to be energetically near the VBM. However, owing to the work function reduction observed in Au nanoparticles~\cite{Zhang_2015}, the Fermi level is closer to the conduction band minimum of WSe$_2$, with Schottky barrier $\Phi_e=0.470\pm0.005 eV$ for electrons (ESB) and $\Phi_h=1.000\pm0.005 eV$ for holes (HSB)~\cite{Smyth_2017}. 

Such suppression results from the nanoscale structure of Au. In these conditions, the nanoparticles have a significantly reduced work function. This is clear from the Au 4f shifts in Fig.~\ref{fig:fig_S2}~a. When 12 Å Au are deposited on WSe$_2$ surface, the 4f$_{7/2}$ peak shifts about -420$\pm$8 meV to higher binding energies, owing to the reduced work function. Upon further increase of the thickness to 36 Å, the film still remains disconnected, but the islands grow, thus getting closer to the bulk work function: the Au 4f shift decreases to -220$\pm$7 meV. Au 4f peak positions are determined by fitting the 4f$_{7/2}$ with a Doniach-Sunjic lineshape, after Shirley background subtraction of the whole 4f doublet. Errors are propagated from the standard deviation on the fit result.   

As the islands are disconnected, a static electric dipole also builds at the interface, that further contributes to the band bending suppression. This is clear from the monotonous positive  shifting of the W 4f peaks in Fig.~\ref{fig:fig_S2}~b: at 12 Å, it is +270$\pm$8 meV, while at 36 Å, it is +440$\pm$9 meV. W 4f peak positions are determined by fitting the 4f$_{7/2}$ with a Voigt lineshape, after Shirley background subtraction of the whole 4f doublet. Errors are propagated from the standard deviation on the fit result.   

It does not appear that Au evaporation produces significant chemical interaction with WSe$_2$, as the core-level lineshapes are remarkably identical to each other, aside from the aforementioned shifts. There are also no significant hints of hybridization in the bandstructure, as explained in the previous section.

Finally, we have one more method to pinpoint the band alignment, i.e. the use of time-resolved ARPES. As it can be readily seen in the overlay of Fig.~\ref{fig:fig_S2}~c, the Fermi edge is indeed about 0.5 eV away from the bottom of the conduction band. More precisely, we find the valence band minimum 612$\pm$2 meV above the Fermi edge. However, we have to consider that at the time delay in which we have sufficient population of the bands (reported in the inset in Fig.~\ref{fig:fig_S2}~c) we also have a transient relative shifting of the charge state of the two materials, that adds up to 140$\pm$5 meV (see Fig.3a). The equilibrium position of the conduction band minimum is therefore 472$\pm$5 meV above the Fermi level, and this constitutes the electron Schottky barrier.

\section{Lifetime of $\Sigma$ valley population}

To perform the fits of the $\Sigma$ valley population, we built a fitting function that consists of a double exponential decay convolved with a gaussian instrument response function.

\begin{figure}[h!]
\centering\includegraphics[width=\textwidth]{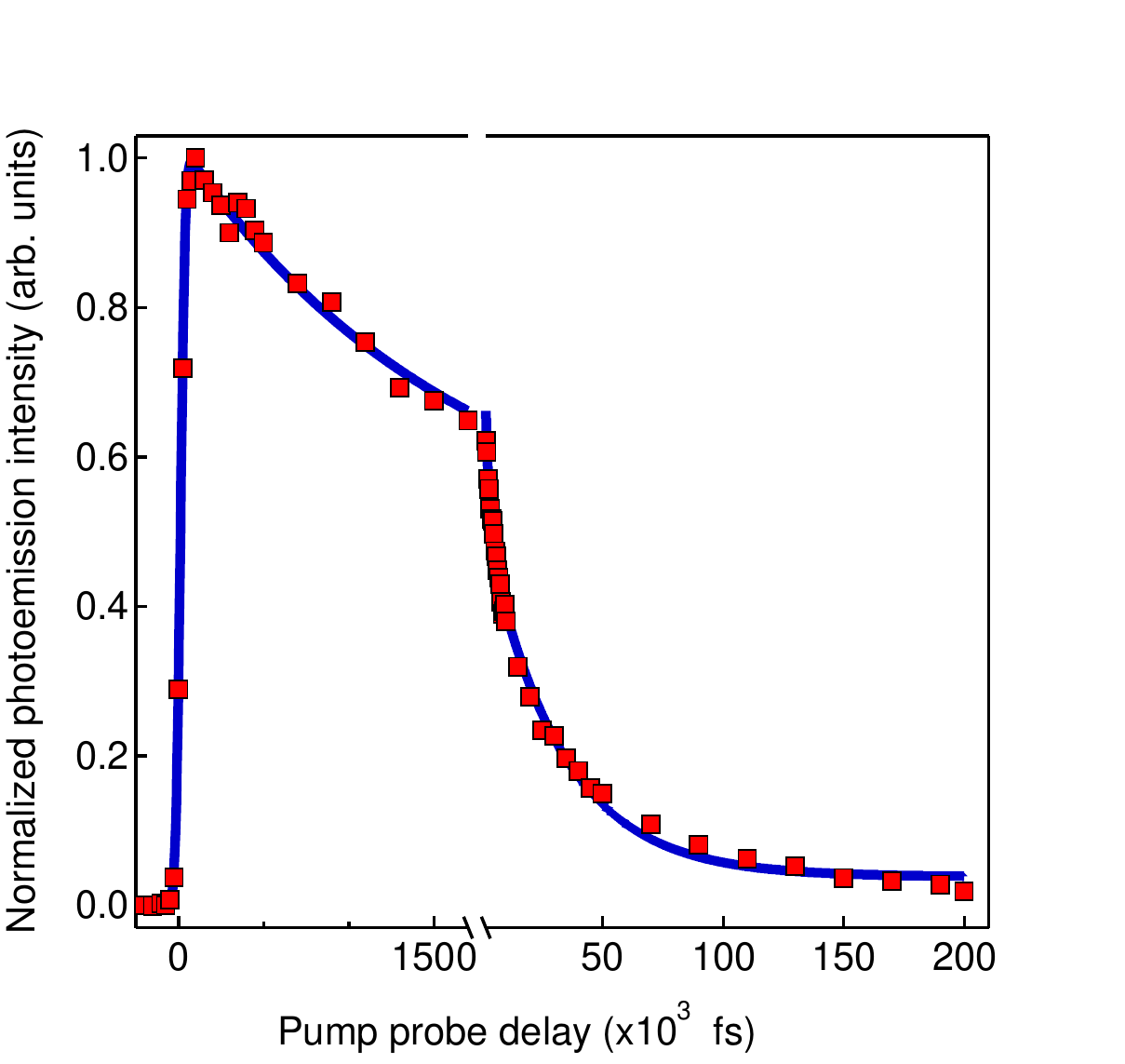}
\caption{\textbf{Long delay $\Sigma$ valley population for pure WSe$_2$.} Combination of data and fit for pure WSe$_2$ measured with 800 nm pump, 0.6 mJ/cm$^2$ fluence. The long delay range shows a two timescales decay.}
\label{fig:fig_S5}
\end{figure}

This allowed us to isolate the intrinsic and interfacing-dependent timescales.We first considered a dataset measured for the WSe$_2$ bare surface in the same experimental conditions, but with a much wider delay range, up to 200 ps. Fitting the population of the $\Sigma$ valley with the convolved double exponential (with fixed gaussian FWHM of 40 fs), returned two timescales: 1502$\pm$78 fs and 31$\pm$2 ps. The result is reported in Fig.~\ref{fig:fig_S5}.

The first time-constant seems rather short with respect to the dark exciton lifetime or defect recombination times reported in literature~\cite{Massicotte_2016}. It thus might be attributable to dark exciton diffusion in the bulk, away from the probing depth of ARPES.

In the dataset discussed in the main text, the delay range is much smaller, so only the short timescale is relevant. It is evident from the data, however, that a second, shorter timescale arises in the heterostructure. We thus considered the curve of Au/WSe$_2$ at low temperature, showing the fastest dynamics.
By fixing the longest timescale to be 1502 fs, we obtain the shortest to be 240$\pm$28 fs. The amplitude of the fast decay is 72$\pm$2\% of the signal. 
These two timescales are sufficient to fit the other two curves in Fig.3c with high reliability by only changing the ratio between them. The one for Au/WSe$_2$ at RT is fitted by reducing the short timescale to the 36$\pm$8\% of the signal, while the pure WSe$_2$ converges with only 2$\pm$10\% of the short timescale, indicating that it is well fitted by the single exponential of 1.5 ps. 

\section{Fluence dependence of $\Sigma$ valley population}

We discuss here the procedure to extract data from the 1030 nm pump dataset. The measurement was performed at fixed delay of +400 fs, i.e. after the ultrafast charge transfer mechanisms have taken place, thus in the condition to observe the exciton population generated in WSe$_2$ by two photon processes. The fluence was changed from 0.28 to 2.34 mJ/cm$^2$.

In these conditions, the electronic temperature is between 1500-2000 K and, while still lower than 5800 K (the minimum electron Schottky barrier, which also increases transiently), this is sufficient to produce a background signal from the Au electrons that needs to be removed in order to isolate the $\Sigma$ valley population. 

\begin{figure}[h!]
\centering\includegraphics[width=\textwidth]{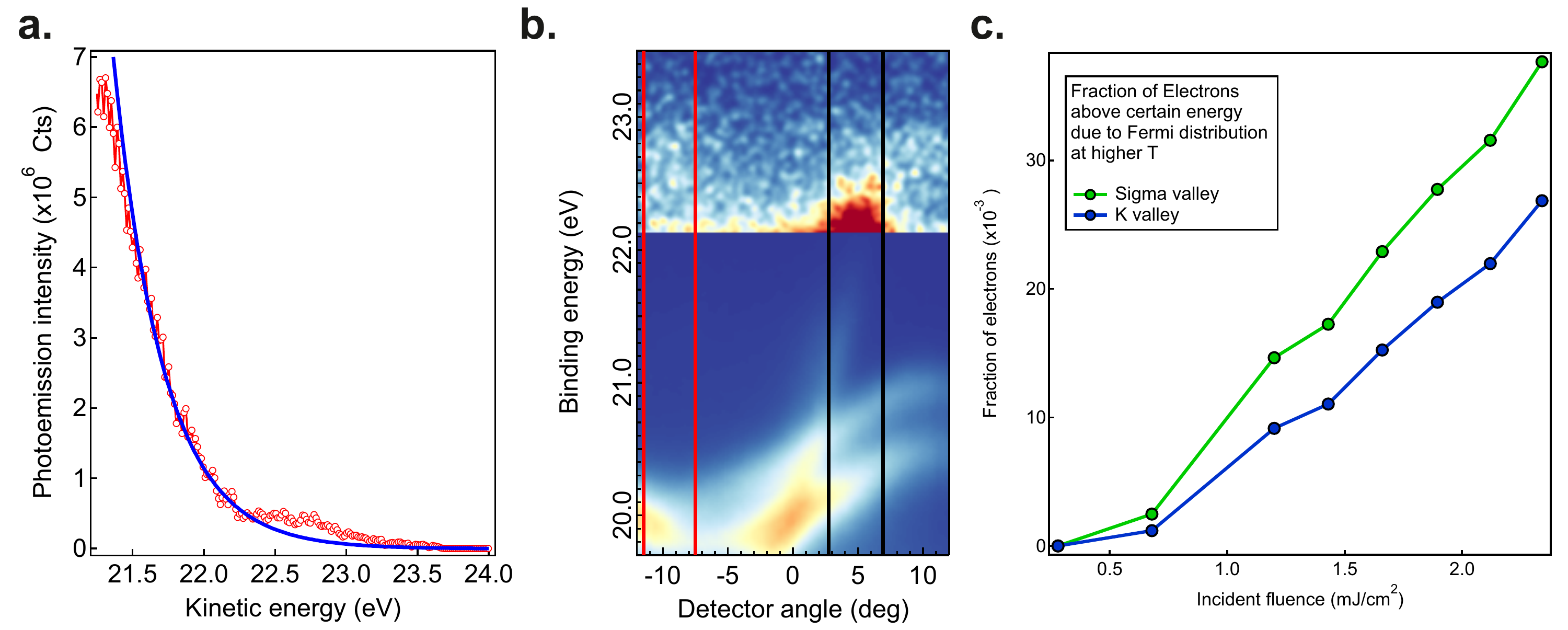}
\caption{\textbf{Analysis of 1030 nm measurements.} a. Typical result of background fitting for the extraction of fluence dependence of $\Sigma$ valley population. The fluence is 1.66 mJ/cm$^2$, the delay +400 fs. The red circles show the data as obtained from tr-ARPES signal integration in the momentum range shown in panel b by the red lines. The blue curve shows the exponential background. b. tr-ARPES map with 1030 nm pumping. The red lines mark the range of integration for the EDC used for $\Sigma$ valley population extraction, while the black lines show the range for Fermi edge fitting. c. Plot of the integral of the Fermi-Dirac distribution between the energy of the $\Sigma$ and K valley and infinity, as a function of the electronic temperatures corresponding to the fluences used in Fig.3d.}
\label{fig:fig_S6}
\end{figure}

The shape of the Au bands in the momentum region corresponding to the WSe$_2$ sigma valley (between the red lines in Fig.~\ref{fig:fig_S6}~b) is such that a fluence dependent fit of the Fermi edge is not reliable at high fluences. For this reason, an exponential decay background was used above the Fermi level. A typical result is shown in Fig.~\ref{fig:fig_S6}~a. However, given the less physically solid choice of fitting function, a routine was employed to improve statistical accuracy: for each point of the fluence dependence, the background fitting and subtraction was repeated six times, with slightly different choices of fitting intervals. The error bar in the data reported in Fig.3d are determined as the standard deviation of the result distribution.

Fermi edge extraction was instead carried out in the momentum range corresponding to the sp-band, marked by black lines in Fig.~\ref{fig:fig_S6}~b. In this region, the Fermi distribution could be fitted reliably at all fluences, and the error reported is the standard deviation obtained by least square fitting.

Finally, we report here a plot of the integral of the Fermi-Dirac electron distribution above the energy levels corresponding to the $\Sigma$ and K valleys. If we exclude the lowest fluence, that produces a negligible increase in the electronic temperature, and fails to produce a detectable population above the WSe$_2$ valence band energy, as the hot electron population scales linearly in this range of fluences. 

\section{Negligibility of Seebeck effect and surface photovoltage}

Regarding the shifts observed in Fig.~3b, we have examined if they result from a combination of surface photovoltage and Seebeck effect (transient voltages due to transient temperature gradients). Both effects can be ruled out. 

On Bare WSe$_2$ the bands shift by less than 7 meV in any condition of fluence, pump wavelength and temperature, thus ruling out any intrinsic effects of the semiconductor. The effect is also too short-lived with respect to classic surface photovoltage.

The Fermi edge of Au does not shift due to temperature changes. This is what is measured  by the Seebeck coefficient, which for Au is below 3 $\mu$V/K up to 2500K. This means that a change of temperature of 3000 K results in a shift of less than 9 meV. We observe shifts of 20 meV (electronic temperature change $\Delta T_e$=700 K, Seebeck shift 2 meV), or 40 meV ($\Delta T_e$=800 K, Seebeck shift 2.4 meV) or finally of 200 meV ($\Delta T_e$=3000 K, Seebeck shift 9 meV).

\section{Energy flow across the interface}

The band alignment suggests that under photoexcitation with photon energies below the Au interband threshold ($<$2 eV) plasmon-generated hot electrons can be injected at much larger rates than hot holes thus dominating the charge transfer mechanism. The still sizable Schottky barrier suppresses the diffusion of thermalized electrons across the interface for low photoexcitation fluences ($<$10 mJ/cm$^2$).

At higher pump photon energies, it becomes possible to excite large populations of deep lying holes with long lifetimes, giving rise to complex energy exchanges that may tip the balance in favour of hot-hole injection~\cite{Dunklin_2019,Dong_2021}, but this falls beyond the scope of our work. We will restrict our discussion to pump photon energies below the A exciton resonance of bulk WSe$_2$ (1.626 eV at room temperature~\cite{Arora_2015}).

A simple electrostatic calculation considering the peak chemical potential shift $\Delta E = 40$ meV and the nanoparticle capacitance $C=$3~aF allows to calculate the net number of injected electrons as $n=\Delta E \cdot C \leq 1$. Combining this with the number of absorbed photons per nanoparticle at 800 nm, considering an absorption of 0.1\% of the nanoparticles (from the FDTD calculations), we get a quantum efficiency (in gap photon to hot electron in WSe$_2$) of about 2\%, in line with other reports on similar systems. We argue, however, that the flow of energy carried by unbalanced charges is smaller than the total energy flow across the interface, owing to the fact that both HSB and ESB are smaller than the photon energy at any pump excitation wavelength.

\section{Probing depth of ARPES}

When considering photoemission data from a heterogeneous sample, it is important to remember that the probing depth of ARPES is extremely short, in the range 10-5 Å at the photon energies employed in the current experiment (21.7 eV). This means that the experiment is predominantly sensitive to the top facet of the Au islands and the open areas of WSe$_2$, while the interface between the islands and the semiconductor is more challenging to access experimentally. 

Our analysis overcomes this hurdle in two ways. Firstly, we focused on observables that do not require the direct observation of the spatial region below the particle. 
The band alignment analysis is supported to a very high degree of consistency by core-level data (more bulk sensitive) and theoretical calculations. The dynamics is explored considering the electronic temperature of Au, its chemical potential, the band positions in WSe$_2$, i.e. properties that can be considered homogeneous at the scale of a single particle or gap (approx. 10 nm) at these timescales ($>$ 10 fs). Secondly, to assess for example population dynamics, we put ourselves in a condition where we are intrinsically selective of the carriers generated by injection by tuning the excitation wavelength.

The latter strategy is viable because the signal does carry information on the WSe$_2$ Bloch eigenfunctions below the islands, at least to a degree. A careful comparison of the MDCs of the bands of bare WSe$_2$ to the same states in the heterostructure, shows that they are broadened to a FWHM of 0.1 Å$^{-1}$. This corresponds, in real space, to approx. 10 nm, i.e. the average gap between the islands, suggesting that, in a three-step picture of photoemission, the confinement of the ARPES wavefunction happens in the final step. The information carried by the angular distribution of the photoelectrons, however, relates to the coherence length of the electron wavepacket in the solid within the plane. In this direction, the inelastic mean free path of the electrons is rather large, and no significant losses are encountered at such energies. Therefore, while the signal arising from below the nanoparticles might be suppressed, the ARPES signal should still carry information regarding the interface.

\section{Non-thermal electrons}

As discussed in \cite{Mueller_2013}, in the femtoseconds immediatley subsequent optical excitation the electronic distribution in the proximity of the Fermi edge assumes a shape that is not described by a Fermi-Dirac distribution. It is instead formed by a Fermi-Dirac distribution of higher temperature, with superimposed steps with exactly the width of the photon energy $\hbar \omega$. A step-wise decrease in the range [$E_f-\hbar \omega$, $E_f$] as electronic states below the Fermi level are depleted, and a step-wise increase in the range [$E_f$, $E_f+\hbar \omega$]. The steps might have a more complicated structure, arising from the details of the DOS of the metal within $\hbar \omega$ from the Fermi level. Aluminum, for example, is theoretically predicted to display a small peak in each step, owing to a local maximum in the DOS just below $E_f$. The DOS of Au, instead, is rather flat until 2 eV below $E_f$, thus producing flat, step-like features.
We investigate the existence of such non-equilibrium distributions for two reasons: to ensure that large non-thermal distributions of electrons do not affect our fitting of the Fermi edge (in particular the energy position), and to understand the role of non-thermal electronic distributions in the early dynamics of the heterostructure.

\begin{figure}[h!]
\centering\includegraphics[width=\textwidth]{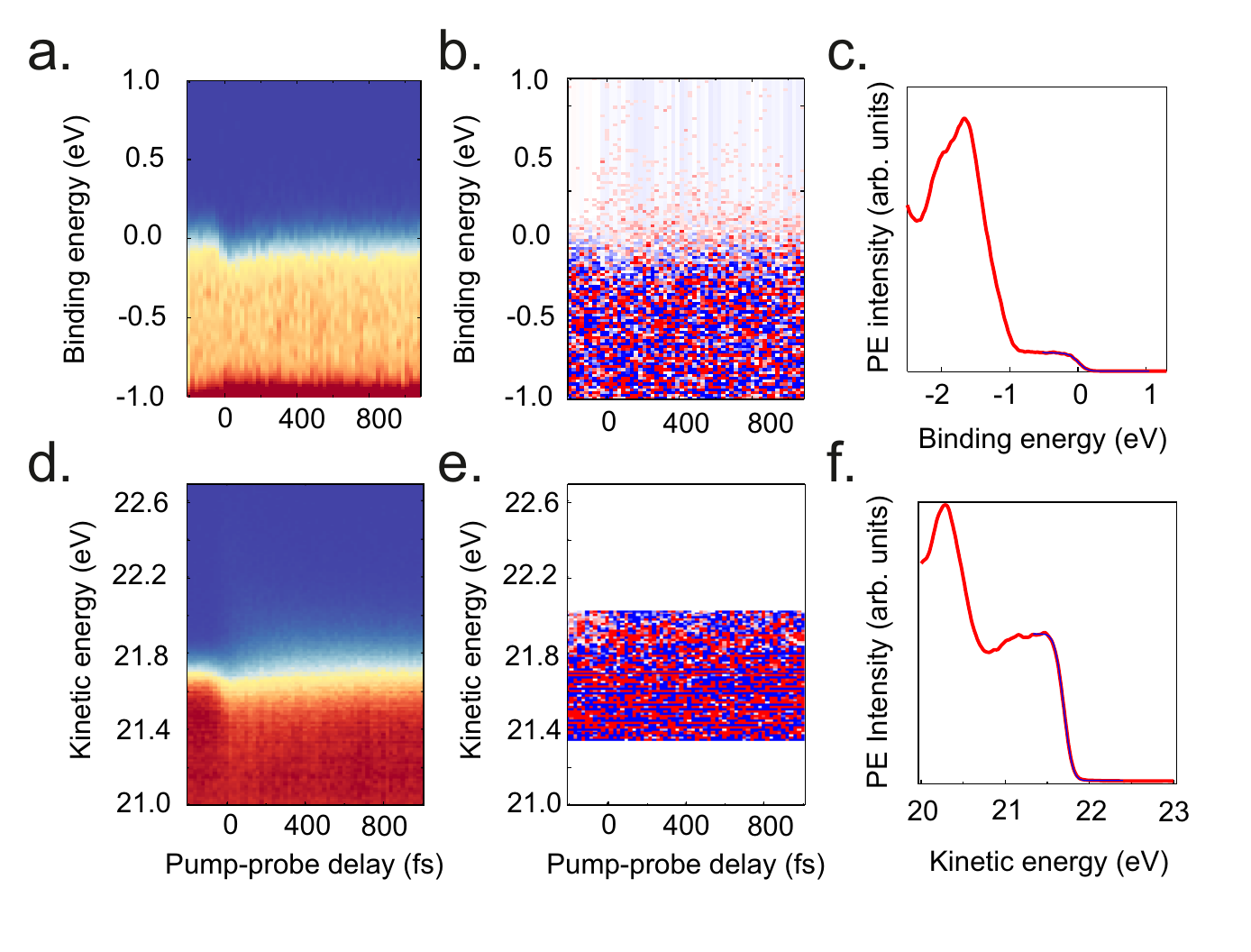}

\caption{\textbf{Fits of the Fermi edge.} a. Surface state Fermi edge dynamics map showing an EDC for every delay. EDCs are extracted integrating k$_x$ [-0.5, 0.5] and k$_y$ [-0.5,0.5] to encompass the Shockley surface state. b. Residuals at each delay, the image colorscale has been scaled to cover $\pm 0.5 \%$ of the Fermi edge amplitude. c. Representative EDC (red, solid line) with Fermi edge fit (blue solid line). d. Fermi edge dynamics map showing an EDC for every delay. EDCs are extracted integrating a range including the sp-band. e. Residuals at each delay, the image colorscale has been scaled to cover $\pm 0.3 \%$ of the Fermi edge amplitude. f. Representative EDC (red, solid line) with Fermi edge fit (blue solid line). }
\label{fig:fig_S7}
\end{figure}

As can be seen in Fig.~\ref{fig:fig_S7}, both the fits of the Fermi edge return no significant non-thermal contributions superimposed on the Fermi-Dirac distribution. The fits have been performed by fitting a Fermi-Dirac distribution convolved with a Gaussian distribution to simulate experimental distribution. The width of the Gaussian was obtained by fitting the negative delays while constraining the temperature to be 70 K, and found to be 150 meV. Then, the Fermi edge was fit leaving the amplitude, the energy position and the temperature as free parameters. The fits of the sp-band have been performed in a rather restricted energy range to avoid contamination of the signal with the intensity of WSe$_2$ conduction band K points.

The lack of obvious non-thermal distributions, or in general of strong trends in the residuals of the fits, indicates that the fitted Fermi edge position discussed in the manuscript is reliable.
More subtle is the interpretation of the role of non-thermal populations in the dynamics of the heterostructure. While we would expect to observe non-equilibrium distributions at these incident fluences and delay ranges, their absence might suggest that all hot carriers are injected in the semiconductor. However, to assess the non-thermal component of the electronic distribution directly, it would be necessary to have a signal-to-noise ratio in the residuals much higher than the one of the current experiment. Such dedicated experiment falls beyond the scope of the current work.

\section{Two temperature model for electron-lattice equilibration in Au}

In Fig.~3a of the main article we compare the experimental results to the prediction of the two temperature model (2TM)\cite{Chen_2006,Ratchford_2017}. A simple two temperature model was set-up to create a reference of bulk Au without charge-transfer, in the same conditions of excitation density.
The equations employed are:
\begin{align}
    \label{eq:eq3}
    (\gamma_e \cdot T_e) \cdot \dfrac{\mathrm{d}T_e}{\mathrm{d}t} = -G_{e-ph}(T_e-T_l) + P(t)\
\end{align}
\begin{align}
    \label{eq:eq4}
    C_l \cdot \dfrac{\mathrm{d}T_l}{\mathrm{d}t}= G_{el-ph}(T_e-T_l)\
\end{align}
with 
\begin{align}
    \label{eq:eq5}
    P(t)=\dfrac{A}{\Delta t / 2 \cdot \sqrt{\pi/\mathrm{ln}(2)}} \cdot \mathrm{e}^{\dfrac{-4 \mathrm{ln}(2)(t-t_0)^2}{\Delta t^2}}\
\end{align}
values of the parameters are $\gamma_e=70$ $\mathrm{J m^{-3} K^{-2}}$, $G_{e-ph}=3 \times 10^6 $ $\mathrm{W m^{-3} K^{-1}}$ at 300 K, $G_{e-ph}=2 \times 10^6 $ $\mathrm{W m^{-3} K^{-1}}$ at 70 K, $C_l=2.4 \times 10^6$ $\mathrm{J K^{-1}}$ at 300 K, $C_l=2 \times 10^6$ $\mathrm{J K^{-1}}$, $A=10^8$ $\mathrm{V m^{-1}}$, $\Delta t= 36$ $\mathrm{fs}$.

\section{Optical properties and thickness of free-standing WSe$_2$}

For the FED experiments we have estimated the thickness of bare WSe$_2$ flakes from their optical properties (absorption spectrum). Subsequently, the thickness of the deposited Au (with electron beam evaporation) was controlled with a quartz crystal microbalance.   

\begin{figure}[h!]
\centering\includegraphics[width=0.8\textwidth]{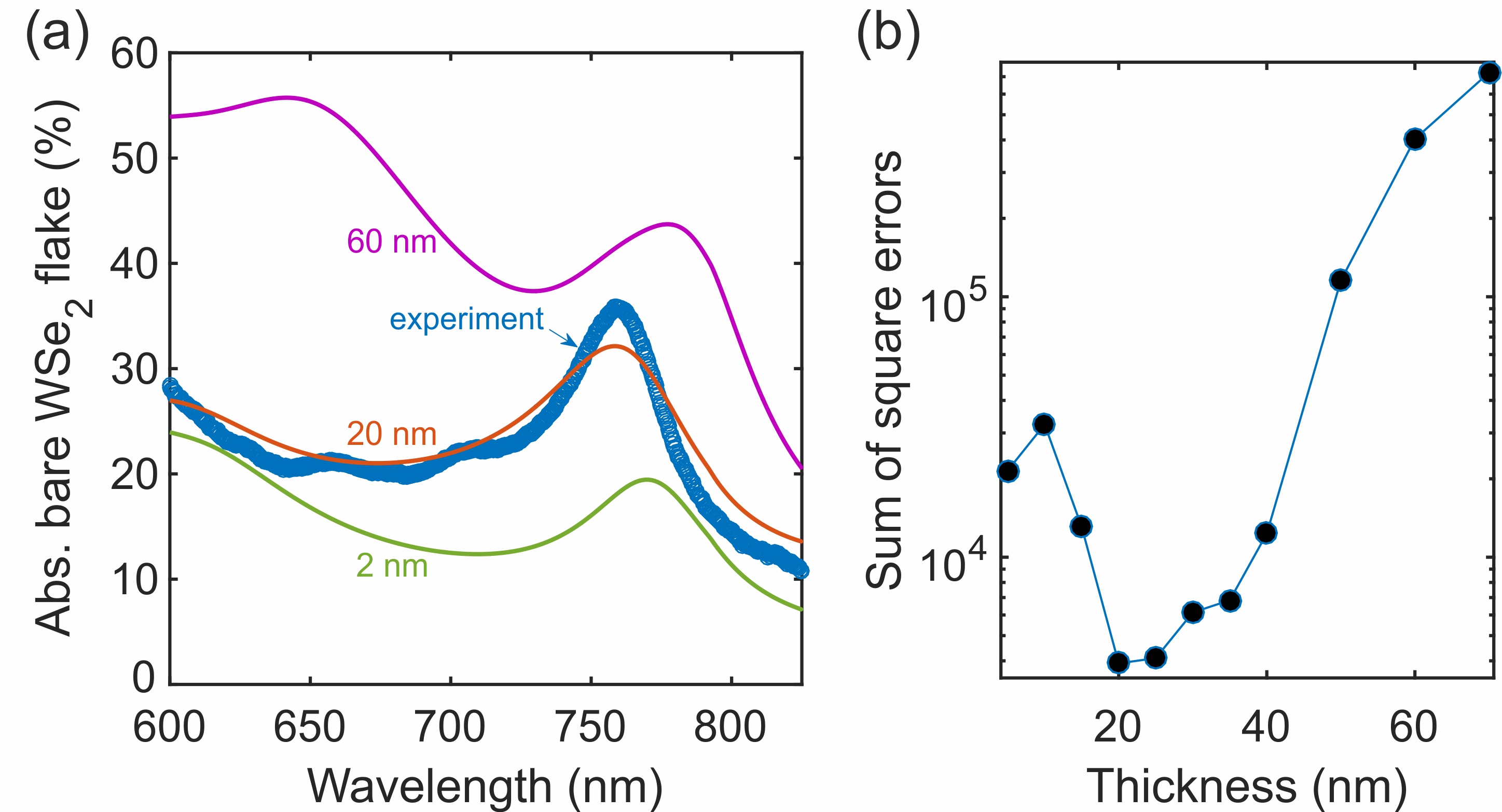}

\caption{\textbf{Fitting the absorption spectrum of bare WSe$_2$ flakes used for FED to extract the thickness.} (a) The experimental absorption spectrum of a bare WSe$_2$ represented with the calculated spectra for 20 nm and 30 nm thickness. (b) The sum of square errors (residuals) as a function of the thickness.}
\label{fig:fig_S8}
\end{figure}

The values for the wavelength-dependent real (n) and imaginary (k) refractive indices of multilayer WSe$_2$ are taken from the work of Gu et al.\cite{Gu_2019}. The absorption spectrum is calculated with the transfer matrix method as a function of the WSe$_2$ thickness. The calculated spectra are compared with the experimental results (Fig.~\ref{fig:fig_S8}~a). The sum of square errors (residuals of the fitting) is minimized for a thickness of 20-25 nm (Fig.~\ref{fig:fig_S8}~b). 

\section{Extraction of MSD and lattice temperatures from FED}

\begin{figure}[h!]
\centering\includegraphics[width=\textwidth]{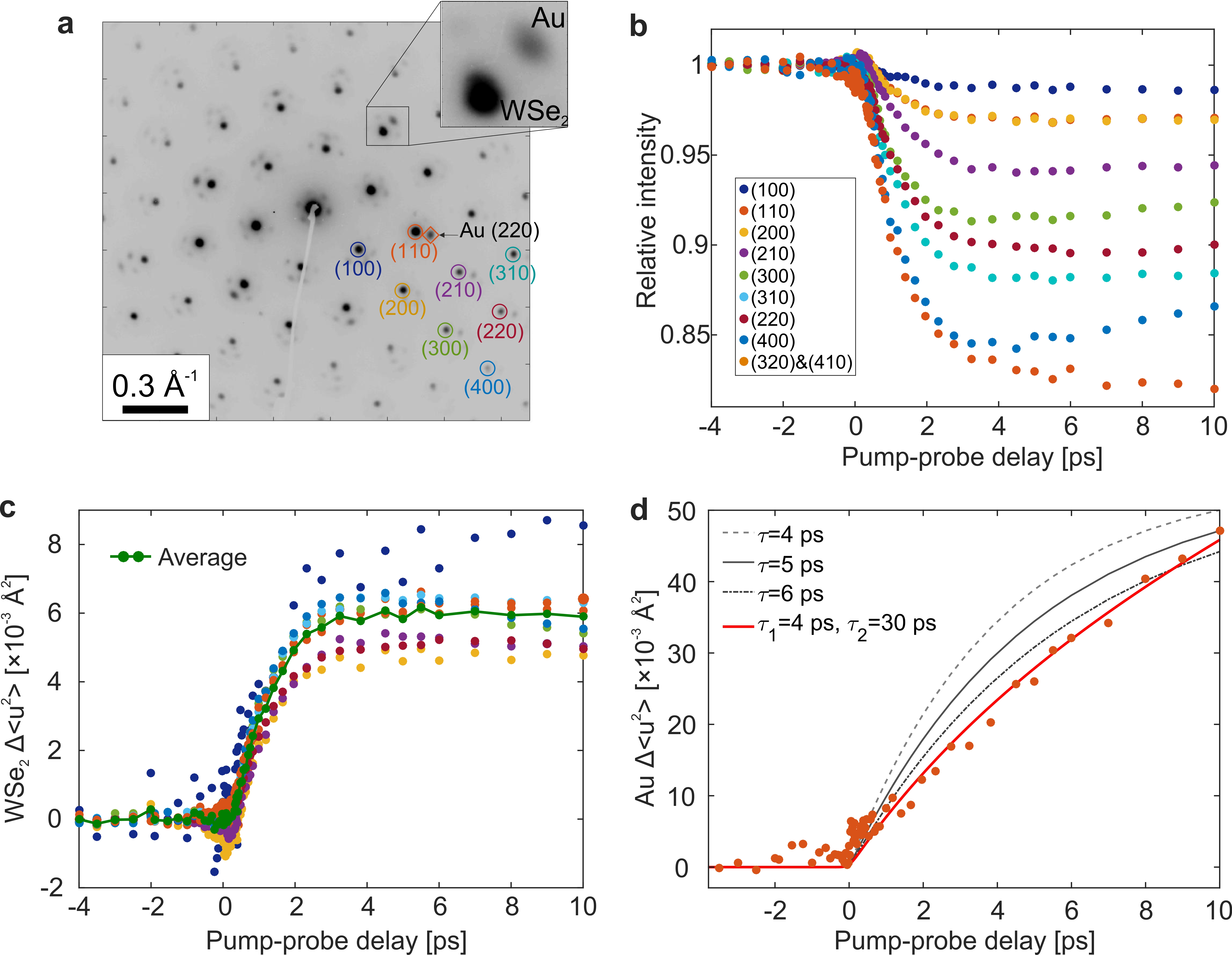}

\caption{\textbf{FED of Au/WSe$_2$ heterostructures pumped at the A-exciton resonance}. (a) The electron diffraction pattern of Au/WSe$_2$ (logarithmic scale). The dark spots represent areas with high intensity of diffracted electrons. The two materials form two hexagonal patterns with strong and weak intensity of diffracted electrons belonging to WSe$_2$ and Au, respectively (see inset). (b) After excitation (Delay>0) all diffraction peaks decay due to the Debye-Waller effect~\cite{Peng,Gao_Peng}. The inset shows the index of each diffraction peak. (c) The intensity decay from each diffraction peak is used to extract the time-dependent atomic MSD of WSe$_2$ (same color code as in (a and b) and the average of all peaks (green dot-line). (d) The time-dependent atomic MSD of Au following the same procedure. The solid lines represent exponential decay functions of various time-constants.}
\label{fig:fig_S9}
\end{figure}

An example of a static diffraction pattern of the Au/WSe$_2$ heterostructures in logarithmic scale is shown in Fig.~\ref{fig:fig_S9}~a. The bright hexagonal pattern corresponds to single-crystalline, multilayer flakes of WSe$_2$. The less intense diffraction pattern corresponds to the epitaxially grown, (111)-oriented, nanoislands of Au (inset of Fig.~\ref{fig:fig_S9}~a). In the time-resolved experiments the lattice dynamics are initiated by femtosecond laser pulse pumping the electrons. The lattice dynamics are probed with ultrashort electron pulses at selected pump-probe delays. For each diffraction peak of WSe$_2$ we extract the relative intensity (Fig.~\ref{fig:fig_S9}~b) and subsequently the change of the atomic MSD (Fig.~\ref{fig:fig_S9}~c). For Au the temporal evolution of the MSD (Fig.~\ref{fig:fig_S9}~d) cannot be described by the single-exponential dynamics measured previously for bulk Au and Au nanoclusters on insulating substrates.  

\begin{figure}[h!]
\centering\includegraphics[width=0.5\textwidth]{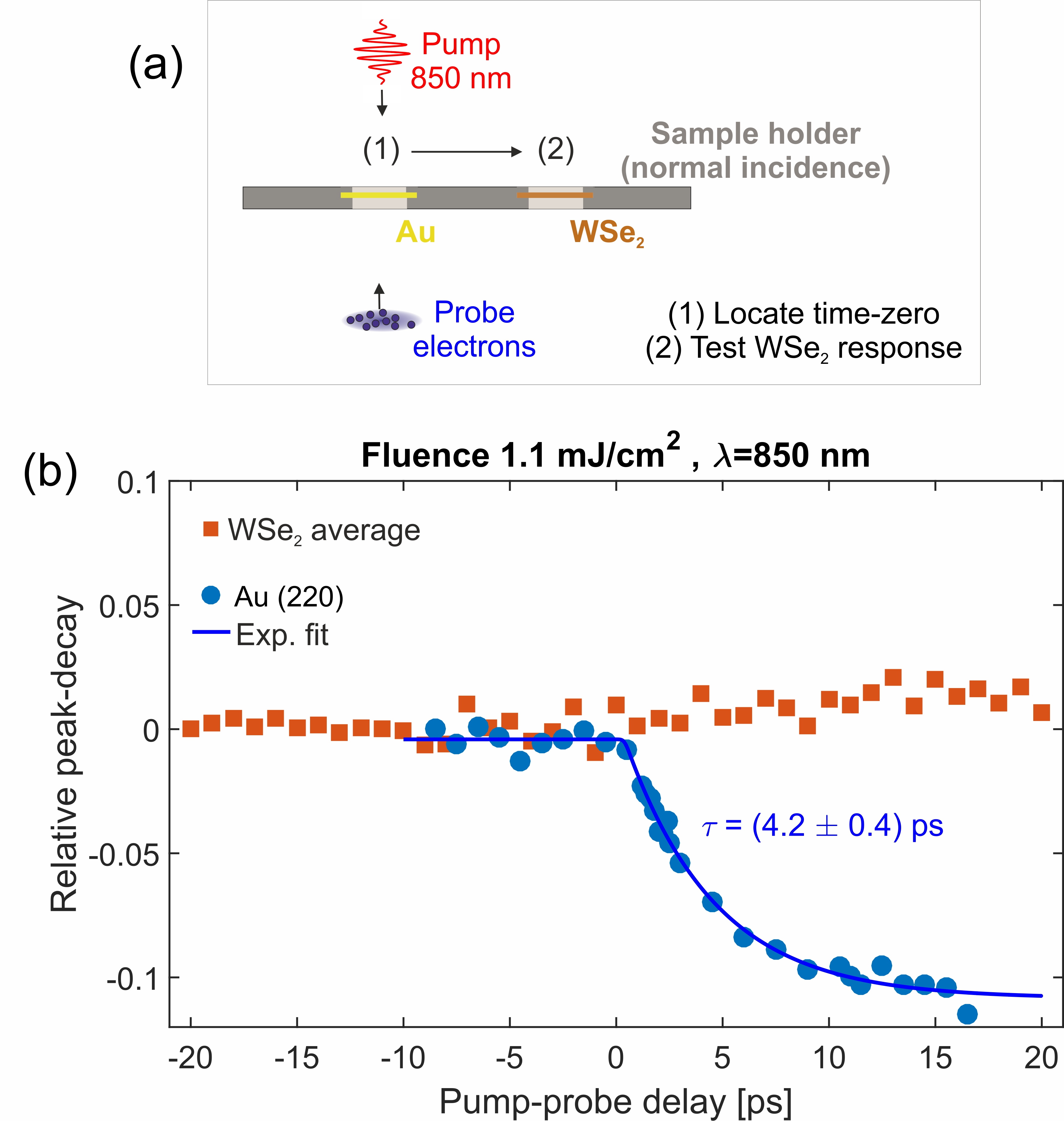}
\caption{\textbf{No lattice response of bare WSe$_2$ flakes exposed to sub-band-gap light.} (a) The experiment used for finding temporal overlap of pump and probe and then testing the lattice dynamics of bare WSe$_2$. (b) The relative peak-decay for the (220) peak of a Au thin-film and for the average of the WSe$_2$ peaks.}
\label{fig:fig_S10}
\end{figure}

To confirm that sub-band-gap light does not induce any measurable lattice dynamics in bare WSe$_2$, we have performed the experiment shown in Fig.~\ref{fig:fig_S10}~a. First, we found spatial and temporal overlap of pump (850 nm) and probe (electrons), and then we moved to a bare WSe$_2$ flake and repeated the FED scan in the $\pm$20 ps range. 

\begin{figure}[h!]
\centering\includegraphics[width=0.8\textwidth]{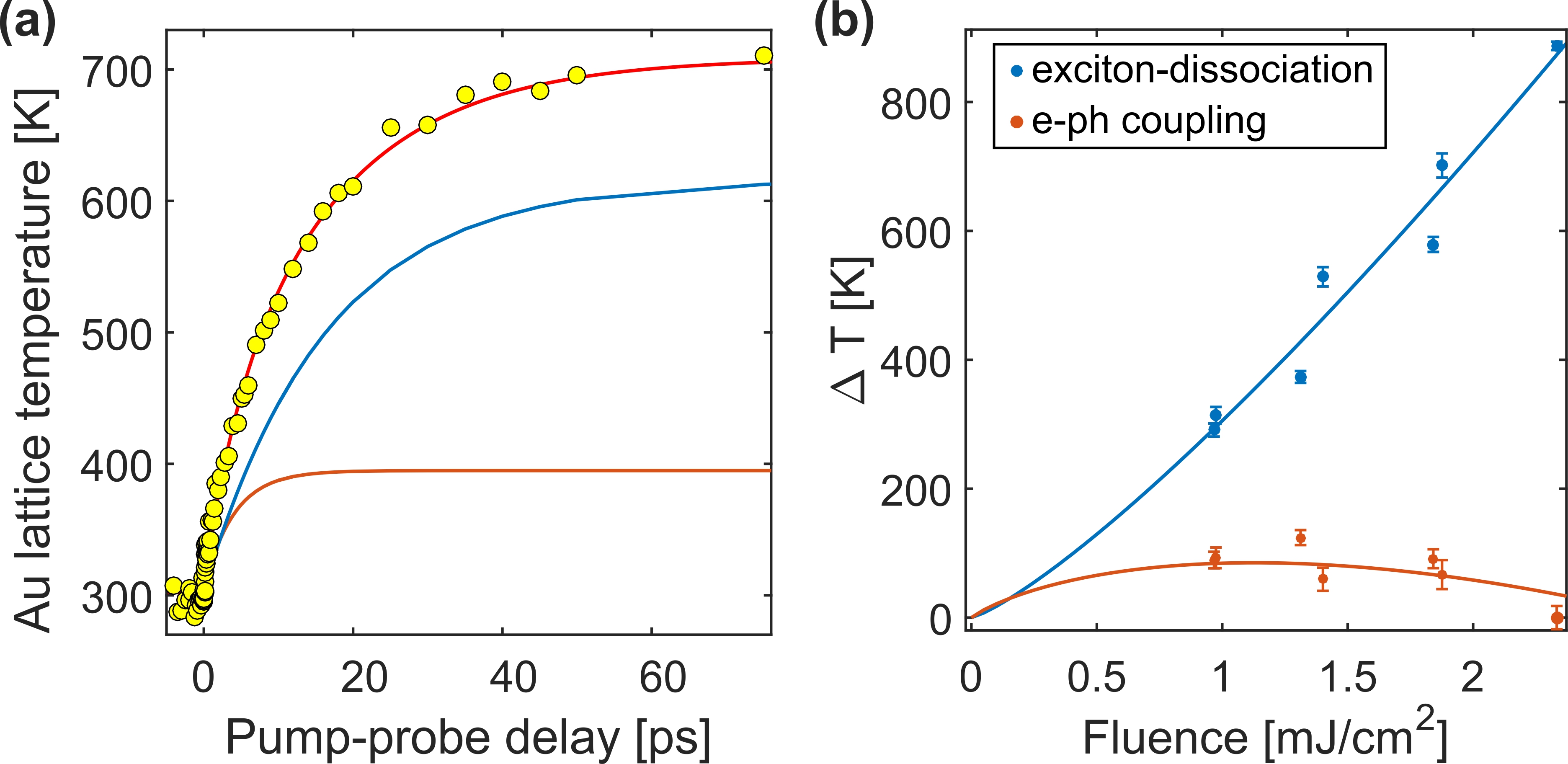}
\caption{\textbf{Lattice temperature evolution of Au nanoislands.} (a) Lattice temperature evolution of Au nanoislands on WSe$_2$ with sub-band-gap light. The experimental data (points) are fitted with a biexponential function (red) with time-constants 4 ps (orange) and 16 ps (blue). (b) The fluence dependent temperature rises of the fast and slow process, which are attributed to e-ph coupling and exciton dissociation, respectively.}
\label{fig:fig_S11}
\end{figure}

In contrast, sub-band-gap light initiates a strong lattice response for Au-decorated WSe$_2$ flakes. An incident laser fluence of 0.97 mJ/cm$^2$ can cause a 410 K temperature rise of the Au nanoislands (Fig.~\ref{fig:fig_S11}~a).The representation of the temperature of evolution of Au with a biexponential function is 20\% more accurate, in terms of the sum of square errors, compared to a single exponential. The two processes have time-constants 4 ps and 16 ps and cause temperature rises of 90 and 310 K, respectively.  

Based on the results and discussions of the main article, the fast process is attributed to electron-phonon coupling in Au and the slow process to exciton dissociation. Fig.~\ref{fig:fig_S11}~b shows the temperature rises caused by the two processes as a function of the incident laser fluence. The solid lines are fittings with functions of the form $aF+bF^{c}$, where $F$ is the fluence. Noticeably, as the fluence increases the heating of Au is dominated by exciton dissociation, while  electron-phonon coupling in Au is suppressed. This observation corroborates our conclusion that plasmons and hot electrons in Au can induce nonlinear absorption and rapid energy transfer into WSe$_2$. 

\medskip

%
\bibliographystyle{MSP}
\bibliography{SI}